\documentstyle[prb,twocolumn,aps,psfig]{revtex}
\tighten
\begin{document}

\title{Charge-density-waves and superconductivity 
as an alternative to phase separation in the infinite-U 
Hubbard-Holstein model}
\author{F. Becca, M. Tarquini, M. Grilli, and C. Di Castro}
\address {Dipartimento di Fisica and Istituto Nazionale per
la Fisica della Materia \\
Universit\`a La Sapienza,\\
Piazzale Aldo Moro, 00185 Roma, Italy}
\maketitle

\begin{abstract}
We investigate the density instabilities present in the 
infinite-U Hubbard-Holstein model both at zero and finite 
momenta as well as the occurrence of Cooper instabilities 
with a specific emphasis on the role of long-range Coulomb 
forces. In carrying out this analysis a special attention 
is devoted to the effects of the strong local $e$-$e$
interaction on the $e$-$ph$ coupling and particularly to both
the static and dynamic screening processes dressing this 
coupling. We also clarify under which conditions 
in strongly correlated electron systems a weak additional
interaction, e.g. a phonon-mediated attraction, can give
rise to a charge instability. In the presence of long-range
Coulomb forces, the frustrated phase separation leads
to the formation of incommensurate charge density waves.
 These instabilities, in turn, lead
to strong residual scattering processes between 
quasiparticles and to superconductivity, thus providing an
interesting clue to the interpretation of the physics
of the copper oxides. 
\end{abstract}


\section{Introduction}
Besides the large critical temperatures, the superconducting
copper oxides display many anomalous normal-state properties
\cite{review}. The understanding of these properties is not
only a fascinating theoretical challenge, but would 
also shed light on the pairing mechanism leading to high
temperature superconductivity. 

The anomalous properties of the normal phase have been 
interpreted along two distinct theoretical lines. The
low dimensionality of these highly anisotropic systems 
and their correlated nature have been proposed to be
 at the origin of a breakdown of the Fermi liquid (FL). 
In particular the concept of Luttinger liquid in two 
dimension \cite{anderson}  was put forward as a new paradigm 
for the normal state of the copper oxides and it was intensively 
investigated \cite{sanseb}.
However, it was recently shown \cite{CDM} that the Luttinger
liquid is only stable in one dimension. Above one dimension,
the Fermi liquid picture is recovered when the bare 
electron-electron ($e$-$e$) interaction is non singular. 
This result would support the alternative aptitude, which 
has been to accept the Landau theory of
normal FL's as a suitable starting point. The anomalous 
properties would then arise as a consequence of 
singular scattering processes at low
energy between the quasiparticles. Along this line 
magnetic scattering has been considered to be 
responsible for both the anomalous  properties of the normal
phase and for the superconducting pairing \cite{pines}. 
 It was also proposed that excitonic scattering
could give rise to the so called marginal FL \cite{varma},
and provide a pairing mechanism. Singular scattering is also 
obtained by gauge fields \cite{nagaosa}, which
arise by implementing the resonating-valence-bond idea in the
t-J model.

The above theoretical lines have more recently been joined
by a different scenario suggesting phase separation (PS) as a
possible source of anomalous scattering and, therefore, 
of anomalous normal-state behavior\cite{emerykivelson,CDGPRL}.
Emery and Kivelson \cite{emerykivelson} suggested that,
although long-range Coulomb (LRC)
forces spoil PS as a static thermodynamic
phenomenon, the frustrated tendency towards PS
may still be important and give rise
to large-amplitude collective density fluctuations.
Approaching the problem within a coarse-grained model,
they suggested that these
fluctuations may be responsible for the anomalous
behaviour of the normal phase and for the superconducting
pairing. In a recent work \cite{CDGPRL}, two of us assessed
 the relevance of charge instabilities
(PS or charge density waves) as a mechanism for anomalous 
scattering, by determining the 
dynamical effective scattering interactions
among Fermi-liquid quasiparticles close to a charge instability,
both in the presence and in the absence of LRC forces.
This analysis consisted in a microscopic treatment
of the Hubbard-Holstein model in the infinite-U limit, finding 
 that, both in the presence and in the absence of
LRC forces, the dynamic effective interaction
have a  singular behaviour, strongly affecting 
the single-particle and the transport scattering time.
This scenario is obviously sensible if i) the considered
microscopic model displays PS for some parameter region and ii)
the real copper-oxide systems actually are in the proximity
of a charge instability \cite{erice}. As far as point i) 
is concerned, PS seems to be a rather generic and robust
phenomenon in the context of strongly interacting systems
\cite{DGpraga}. Indeed, after PS was shown to be present 
in the phase diagram of the t-J 
model\cite{marder,emery,dagotto}, it was pointed out
that PS commonly occurs in models with short
range interaction\cite{CGK,CCCDGR,GRCDK1,RCGBK,CDG,DGpraga,GC}
 provided the strong local 
$e$-$e$ repulsion inhibits the stabilizing
role of the kinetic energy. Moreover it was repeatedly 
claimed that PS and superconductivity can be related 
phenomena irrespective of the
nature of the short-range interaction \cite{DGpraga}.

On the other hand, the frequent occurrence of PS in models
of strongly interacting electrons is made intriguing by
the observation of PS in 
oxygen-doped superconducting copper oxides (${\mathrm La_2CuO_{4+y}}$)
of the 214 class \cite{erice}
Although the electronic origin of PS in these 
cuprates is still to be established, the contemporary presence 
of a strong $e$-$e$ interaction and of PS in a real system and the
robustness of the PS concept in theoretical models
is suggestive. The reason why only ${\mathrm La_2CuO_{4+y}}$ seems
to phase separate is that LRC forces
effectively oppose the separation of charged particles.
Only when the negatively charged oxygen ions are sufficiently mobile
the positive holes can separate being accompanied by
the oxygen countercharges which compensate for the 
charge unbalance. Nevertheless, even in those systems where LRC forces
are present to prevent a thermodynamic instability, phase separation may
remain in the system in the form of a tendency 
toward charge aggregation, possibly giving rise to
superconductivity \cite{emerykivelson,CCCDGR,GRCDK1,RCGBK,CDG,DGpraga,GC}
 or to anomalous normal properties \cite{emerykivelson,CDGPRL}.
In particular it might well happen that the long-wavelength 
density fluctuations associated to PS
are suppressed in favor of shorter-wavelength density fluctuations
giving rise either to dynamical slow density modes \cite{emerykivelson}
or to incommensurate charge density waves (CDW) \cite{notaRCGBK}.
This latter possibility was recently put forward to explain 
neutron-scattering results in a ${\mathrm {La_{1.48}Nd_{0.4}Sr_{0.12}
CuO_4}}$ sample \cite{tranquada}. In this case it was proposed that the 
(low-temperature tetragonal) lattice structure and the filling
(close to 1+1/8 holes per ${\mathrm {CuO_2}}$ cell) were 
suited to pin the 
density fluctuations giving rise to a static CDW phase. 
The formation of striped patterns in the ${\mathrm {CuO_2}}$
planes of Bi2212 compounds was also shown from EXAFS
experiments\cite{bianconi}
Local density fluctuations could also account for 
some results of neutron scattering experiments in 123 materials 
\cite{mesot,tranqrev}

Based on the observation of an antiferromagnetic phase close
to the superconducting one in the phase diagram of the copper 
oxides, previous analyses put emphasis on the role of magnetic coupling 
in originating the slow density fluctuations \cite{emerykivelson}.
However the generic occurrence of PS in theoretical models
with different interactions, indicates that a definite choice
of the mechanism leading to PS could be misleading or premature
in the absence of more stringent experimental indications.
Moreover, while in the models mentioned above the additional
interactions inducing PS are of 
purely electronic origin, it was shown in Ref. \cite{GC}
that, in the presence of a strong local repulsion, 
also the lattice may introduce an effective attraction
determining PS in the Fermi liquid.
This latter result showed within an infinite-$U$ 
three-band Hubbard
model that the instability occurred for reasonable 
values of the $e$-$ph$ coupling indicating that PS
by no means requires unlikely parameters, 
unusual mechanisms or purely electronic 
interactions, but it can simply result from the interplay between
the strong local repulsion and the (weak) additional attraction
provided by the lattice. 
This theoretical observation is accompanied by
some experimental evidences
that the lattice can play a non-negligible role in determining
the physics of the superconducting cuprates \cite{stanford}.
In particular a sizable coupling between some phonons and the 
carriers is implied by  the presence of polaronic
effects \cite{polaronexper} for the very lightly doped compounds,
 by the copper and oxygen 
isotopic effect present in ${\mathrm La_{2-x}Sr_xCuO_4}$,
 by the Fano line shapes
in Raman spectra and by the rather large frequency shift
and linewidth broadening of some phonons at $T_c$.

In the present paper we pursue the investigation along the 
route opened in Ref.\cite{GC} The occurrence of 
a phase-separation instability was justifyied within a
general Fermi liquid analysis demostrating that
the strong interaction is responsible
for vertex corrections, which are strongly dependent on the 
$v_Fq/\omega$ ratio, where $v_F$ is the Fermi velocity and $q$ and 
$\omega$ are the transferred momentum and frequency respectively.
These corrections generically lead to  a strong suppression of the
effective coupling between quasiparticles mediated by (a single)
phonon exchange in the  
$v_Fq/\omega \gg 1$ limit. However, such effect
is not present when $v_Fq/\omega \ll 1$, which is the relevant limit
for the effective interactions entering the Fermi liquid expression 
for the compressibility. In these effective interactions in
the dynamical limit, the $e$-$ph$ coupling is therefore not 
effectively screened opening the way to a possible violation of the
stability criterion for the Fermi liquid in some regions
of the parameter space.
In Ref.\cite{GC}
a detailed analysis was then carried out using a slave-boson
approach for the infinite-U three-band Hubbard model describing the
basic structure 
of a ${\mathrm{CuO_2}}$ plane in copper oxides. In the presence
of a coupling between the local hole density and a dispersionless
optical phonon, it was explicitly confirmed the strong dependence of the
hole-phonon coupling on the transferred momentum versus frequency
ratio and it was also found that the exchange of phonons
leads to an unstable phase with negative compressibility   already
at rather small values of the bare hole-phonon coupling. Close to the
unstable region, Cooper instabilities both in $s$- and
$d$-wave channels  were detected
supporting a possible connection between phase
separation and superconductivity in strongly correlated systems.

We now start from 
the infinite-U single-band Hubbard model
in the presence of an optical phonon coupled to the
local electron density \cite{holstein}. 
Due to its relative simplicity with respect to the three-band
Hubbard model {\it we will be able to extend
the model in a rather direct and straigthforward way so as to
include the LRC forces between the electrons}.
This extension is particularly important since,
as mentioned above, LRC
forces obviously affect the occurrence of PS instabilities
and could provide a clue in explaining the relative rarity of
this phenomenon in the real materials. In this way, as briefly
reported in Ref. \cite{CDGPRL}, we also provide a microscopic
derivation of an incommensurate CDW instability directly from
a system of strongly correlated electrons with all the
physical implications indicated above. Therefore this topic
represents a key issue of our investigation and may definitely
be considered as the main point of our analysis.

In Section II we introduce the model and the formalism. Readers, 
who are not interested in technical details can directly
move to Section III, where we present the results concerning the physical
properties of the model in the absence of LRC forces. The effects of LRC
interactions are reported in Section IV, which thus represents
the core of the present paper, while in Section 
V we discuss the results and we draw our conclusions.

\section{The Hubbard-Holstein model}
Our starting point Hamiltonian is the twodimensional Hubbard model with an
additional dispersionless phonon mode $A$ coupled {\it \`a la } Holstein
\begin{eqnarray}
H & = & -t \sum_{\langle i,j \rangle , \sigma} 
\left( c^\dagger_{i\sigma} c_{j\sigma} + H.c.\right) 
-t' \sum_{\langle \langle i,j \rangle \rangle , \sigma} 
\left( c^\dagger_{i\sigma} c_{j\sigma} + H.c.\right)\nonumber \\ 
& - &\mu_0\sum_{i\sigma} n_{i\sigma}
  +  U\sum_i n_{i\uparrow}n_{i\downarrow} \nonumber \\
& + & \omega_0 \sum_i A^\dagger_i A_i + g \sum_{i,\sigma}
\left( A^\dagger_i+A_i\right) \left( n_{i\sigma} -\langle n_{i\sigma} \rangle 
\right) ,\label{HHHam}
\end{eqnarray}
where $\langle i,j \rangle$ and $\langle \langle i,j \rangle\rangle$ 
indicate nearest-neighbor and next-nearest-neighbor sites respectively
and $n_{i\sigma}=c^\dagger_{i\sigma} c_{i\sigma} $ is the local
electron density.
Since we are interested in the limit of strong local repulsion
we take the limit $U\to \infty$, which gives rise to the local
constraint of no double occupation $\sum_\sigma n_{i\sigma} \le 1$.
To implement this constraint we use a standard slave-boson technique
\cite{barnes}-\cite{slabos} by performing the usual substitution
$c^{\dagger}_{i \sigma}\rightarrow c^{\dagger}_{i
\sigma}b_i, \,\,\, {c}_{i \sigma}\rightarrow b^{\dagger}_i c_{i
\sigma}$. We also use a large-$N$ 
expansion \cite{coleman} in order to introduce a small
parameter allowing for a systematic perturbative expansion
without any assumption on the smallness of any physical quantity.
Within the large-$N$ scheme, the spin index
runs from 1 to $N$ and the constraint assumes the form
$\sum_\sigma c^{\dagger}_{i\sigma} c_{i\sigma} +b^{\dagger}_ib_i = 
{N \over 2}$. A suitable rescaling of the hoppings 
$t \rightarrow t/N$ and $t' \rightarrow t'/N$ must, in this
model, be joined by the similar rescaling of the $e$-$ph$ 
coupling $g \rightarrow {g/ {\sqrt{N}}}$ in order to
compensate for the presence of $N$ fermionic degrees of freedom.
The model can then be represented as a functional integral 
\begin{eqnarray}
 Z & = & \int Dc^{\dagger}_{\sigma}Dc_{\sigma} Db^{\dagger}Db
D\lambda DA DA^\dagger e^{-\int_0^\beta Sd\tau}, \label{funcint}\\
  S & = & \sum_i \left[ \sum_{\sigma} c^{\dagger}_{i \sigma}
 {{\partial c_{i\sigma}} \over {\partial \tau}} 
 +b^{\dagger}_{i} {{\partial b_{i}} \over {\partial \tau}}
+A^{\dagger}_{i} {{\partial A_{i}} \over 
{\partial \tau}} \right] \nonumber \\
& + & \sum_i \left[ i\lambda_i \left( b^{\dagger}_i b_i
-{N\over 2}\right)  \right] +H,  \label{action}\\
 H & = & \sum_{i,\sigma} c^{\dagger}_{i \sigma} c_{i \sigma}
\left( -\mu_0 +i\lambda_i\right)
- {t \over N} \sum_{\langle i j \rangle , \sigma}
\left[ c^\dagger_{i\sigma}c_{j \sigma} b^{\dagger}_j b_i
 + c.c.\right] \nonumber \\
& - & {t' \over N} \sum_{\langle \langle i j \rangle \rangle , \sigma}
\left[ c^\dagger_{i\sigma}c_{j \sigma} b^{\dagger}_j b_i
 + c.c.\right] \nonumber  \\
& -& {g \over {\sqrt{N}}}\sum_{i, \sigma} \left( A_i + A_i^\dagger
\right) \left( n_{i\sigma}-\langle n_{i\sigma} \rangle \right)+
\omega_0\sum_iA^\dagger_iA_i.
\label{ham2}
\end{eqnarray}
where a local Lagrange multiplier field $\lambda_i$ has been
introduced to implement the local constraint.

 At the mean-field ($N=\infty$) level, the model 
of Eqs.(\ref{funcint})-({\ref{ham2}) is 
equivalent to the  standard, purely
electronic $U=\infty$ Hubbard model without coupling to the
phonons, which has been  widely considered in the 
literature\cite{KL}. In fact, at mean field level no
role is played by the phonons because our electron-lattice coupling 
depends on the difference between the local  and the average
density and this difference naturally vanishes in the mean-field 
approximation \cite{notaeqil}. The average number of particle per
cell is $n= (1-\delta)N/2$ and $\delta=0$ corresponds to 
half-filling, when one half electron per cell and per spin 
flavor is present in the system. 

The mean-field self-consistency equations
are obtained by requiring the stationarity of the mean-field
free energy and they determine the values of $b_0^2\equiv N r_0^2=
\langle b_i \rangle^2$ 
and of $\lambda_0\equiv \langle \lambda_i \rangle$. Then the
mean-field Hamiltonian reads
\begin{eqnarray}
H_{MF} & = &  \sum_{k \sigma } E_kc^{\dagger}_{k \sigma } c_{k\sigma}-
(\mu_0 -\lambda_0) \sum_{k \sigma }
c^{\dagger}_{k \sigma } c_{k\sigma} \nonumber \\
 & + & N\lambda_0\left( r_0^2-{1 \over 2} \right)
\end{eqnarray}
where $E_k= -2t r_0^2 \varepsilon_k $ is the quasiparticle band
with $\varepsilon_k \equiv \left(\cos k_x +\cos k_y \right)
+ \alpha \left(\cos (k_x+k_y) + \cos (k_x-k_y) \right)$ 
(we define $\alpha \equiv t'/t$). In particular it
turns out that  
the square of the mean-field value of the
slave-boson field $b_0$, $b_0^2=Nr_0^2=N\delta /2$,
 multiplicatively reduces the 
hopping, $t \to tb_0^2$, thus enhancing the effective mass 
of the quasiparticles. Moreover, at this
level  the single particle self-energy does not introduce a finite
quasiparticle lifetime. Then, in this
model the single-particle Green function of the physical  fermions
at $N=2$ has a quasiparticle pole with a finite residue given by the
square of the  mean-field value of the slave-boson field $b_0^2$.
Thus for any finite doping $\delta$ the system is a T=0 Fermi-liquid.

On the other hand, at half-filling $b_0=\delta=0$ and 
the system is insulating
with a vanishing value of both the quasiparticle bandwidth
(i.e. an infinite quasiparticle effective mass $m^*$) and a vanishing
residuum of the polar part in the single-particle Green function.

As far as $\lambda_0$ is concerned, this quantity rigidly shifts 
in a doping dependent way the bare chemical potential $\mu_0$ and
is self-consistently determined by the following equation
\begin{eqnarray}
\lambda_0 & \equiv &  \lambda_0^0+\alpha \lambda_0^1
= 2t \sum_k f(E_k)\varepsilon_k \nonumber \\
& = & 2t \sum_k f(E_k) \left( \beta_k +\alpha \gamma_k  \right)
\label{lambda}
\end{eqnarray}
where $f(E)$ is the Fermi function and 
$\beta_k\equiv \cos k_x+\cos k_y$ and 
$\gamma_k \equiv \cos (k_x+k_y) + \cos (k_x - k_y)$.

The presence of the coupling with the phonons introduces
new physical effects when one considers the fluctuations
of the bosonic fields. Since only a particular combination 
$a=(A^{\dag}+A)/(2\sqrt{N})$ of the phonon fields $A$ and
$A^{\dag}$ is coupled to the fermions, it is more natural to
use the field $a$ and to integrate out the orthogonal
combination ${\widetilde{a}}=(A-A^{\dag})/(2\sqrt{N})$.
Then the quadratic action for the boson field $a$ reads  
\begin{equation}
H_{\rm {phon}}=N\sum_{n,i} 
{{\omega_n^2+\omega_0^2}  \over {\omega_0}}a_i^\dagger
a_i\,\,, \label{aaction}
\end{equation}
where we have transformed the imaginary time into Matsubara
frequencies. Moreover, it is convenient to work in the 
radial gauge \cite{RN}, the
phase of the field  $b_i=\sqrt{N}r_i\exp(-i\phi) $ is gauged away
and only the modulus field $r_i$ is kept, while $\lambda_i$
acquires a time dependence $\lambda_i\to \lambda_i+\partial_\tau
\phi_i$.
Thus one can define a three-component field  ${\cal A}^{\mu}=(\delta
r,\,\,\, \delta \lambda, \,\,\, a)$ where  the time- and 
space-dependent components are  the fluctuating part of the boson
fields $r_i = r_0 \left( 1+\delta r_i   \right)$, $\lambda_i =
-i\lambda_0 + \delta \lambda_i$ and $a_i$. 

Writing the Hamiltonian of
coupled fermions and bosons as $H=H_{MF}+H_{\rm {bos}}+H_{\rm
{int}}$, where $H_{MF}$ is the above mean-field Hamiltonian,
which is quadratic in
the fermionic fields, $H_{\rm{bos}}$ is the purely bosonic part, also
including the terms with the $a$, $r$ and $\lambda$ bosons appearing
in the action (\ref{action}) and in $H_{\rm {phon}}$,
Eq.(\ref{aaction}). $H_{\rm{int}}$  contains the fermion-boson
interaction terms. 
The single-band $U=\infty$ Hubbard model also contains a four-leg
vertex arising from the hopping part of the Hamiltonian
(see the second term on the r.h.s. of Eq.(\ref{ham2})). 
The two fermionic legs of this vertex can be contracted
(see Fig. 1) giving rise to a leading order self-energy contribution
to the quadratic part of the bosonic Hamiltonian
\begin{eqnarray}
\Sigma(q)  &  =  & -2Nr_0^2t\sum_k \varepsilon_{k+q} f(E_k)  \nonumber \\
           &  =  & 
-{Nr_0^2\over 2} \left[\lambda_0^0 \beta_q
+\alpha \lambda_0^1 \gamma_q \right]
\label{selfenergy}
\end{eqnarray}
Fourier transforming to the
momentum space, the bosonic part of the action reads  
$$
H_{\rm {bos}}=N \sum_{q \mu \nu} {\cal A}^{\mu}(q)B^{\mu \nu}(q) 
{\cal A}^{\nu}(-q)
$$
without explicitly indicating the frequency dependence for the
sake of simplicity and where $\mu,\nu =r,\lambda ,a$.
\begin{figure}
{{\psfig{figure=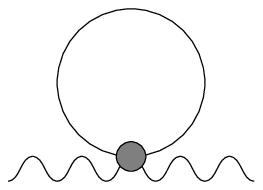,width=3.5cm,angle=-90}}}
{\small Fig 1. Leading-order self-energy contribution to the boson 
propagator from the four-leg vertex in $H_{\rm int}$}
\end{figure}
The matrix $B^{\mu,\nu}$, can be explicitely
determined from Eqs.(\ref{action})-(\ref{selfenergy}) and it is 
found that all elements are
zero except for $B^{r,r}=r_0^2 \left[ \lambda_0^0
\left(1-(1/2)\beta_q \right) +\alpha \lambda_0^1
\left(1-(1/2)\gamma_q \right) \right]$,
$B^{r,\lambda}=B^{\lambda,r}
=ir_0^2$, $B^{a,a}=\left( \omega_n^2+\omega_0^2
\right)/\omega_0 $. 

The last ingredients of our perturbation theory are the vertices
coupling the quasiparticles to the bosons
\begin{eqnarray}
\Lambda^r(k,q)       & = & -2tr_0^2 \left(\varepsilon_{k+q/2}
+\varepsilon_{k-q/2} \right) \label{lamr} \\
\Lambda^\lambda (k,q)& = & i \label{laml}  \\
\Lambda^a (k,q)      & = & -2g \label{lamg}
\end{eqnarray}
allowing to write the interaction part of the Hamiltonian in the 
following form
\begin{equation}
H_{\rm {int}} = 
\sum_{k,q,\sigma}c^{\dagger}_{k+{q\over 2}\sigma}
\Lambda^\mu\left(k,q\right)c_{k-{q\over 2}
\sigma}{\cal A}^\mu\left( q \right).
\end{equation}
The quasiparticle-boson interactions give rise to self-energy
corrections to the boson propagators, which, at leading order
in 1/$N$, are just fermionic bubbles with insertion of
quasiparticle-boson vertices
\begin{eqnarray}
\Pi^{\mu \nu}(q,\omega_m)=
\sum_{k} {
{f\left( E_{k+{q\over 2}} \right)
-f\left( E_{k-{q\over 2}} \right)} \over
{ E_{k+{q\over 2}} - E_{k-{q\over 2}} -i \omega_m}} \nonumber \\
\times {\Lambda}^\mu\left( k,q \right)
{\Lambda}^\nu\left( k,-q \right). \label{pi}
\end{eqnarray}
Once these self-energy corrections are taken into account,
the boson propagator at leading order assumes the form
\begin{eqnarray}
D^{\mu \nu}(q,\omega_m) & = &
\langle {\cal A}^\mu(q,\omega_m){\cal A}^\nu(-q,-\omega_m)
\rangle \nonumber \\
 & = & N^{-1}(2B+
\Pi(q,\omega_m))^{-1}_{\mu \nu}
\label{bosprop} 
\end{eqnarray}
The factor 2 multiplying the boson matrix B arises from the fact that
the bosonic fields in the presently used radial gauge are real.
 
The above formal scheme allows to calculate the leading-order
expressions of the effective scattering amplitude both in the 
particle-hole channel
\begin{equation}
\Gamma (k,k';q,\omega)= -\sum_{\mu \nu}
 {\Lambda}^{\mu}  \left(k', -q
\right) D^{\mu \nu} \left( q, \omega \right)
 {\Lambda}^{\nu} \left(k, q \right). \label{gamma}
\end{equation}
and in the particle-particle channel
\begin{eqnarray}
& \Gamma^C & (k,k';\omega) =  
 - \sum_{\mu \nu} 
 {\Lambda}^{\mu}\left({k+k' \over 2} ,k'-k\right) \nonumber \\
\times & & D^{\mu \nu} (k-k',\omega)
 {\Lambda}^{\nu}\left(-{k+k' \over 2},k-k'\right) 
\label{cooper}
\end{eqnarray}
It should be noted that the boson propagators are of order 1/N
while the occurrence of a bare fermionic bubble leads to a
spin summation and is therefore associated with a factor  N.
Thus, in this 1/N approach, the quasiparticle scattering
amplitudes are residual interactions of order 1/N. 

The form of the static  density-density correlation function at the
leading order is 
\begin{eqnarray} 
P(q,\omega=0) & = & {1\over N}
\sum_{\sigma \sigma'}\langle n_{\sigma}(q) n_{\sigma'}(-q)\rangle 
\nonumber \\
& = &   P^{0}(q,\omega=0)  
       + \sum_{\mu \nu} \chi^{0}_{n \mu}(q,\omega=0) \nonumber \\
& \times & D^{\mu \nu}(q,\omega=0) 
\chi^{0}_{\nu n}(q,\omega=0) \,\,\,\,\,\,\,\,\,
\label{pab}
\end{eqnarray}
where
\begin{equation}
P^{0}(q,\omega) = {1\over N}
\sum_{\sigma \sigma'}\langle n_{\sigma}(q) n_{\sigma'}(-q)\rangle_{0}
\label{p0ab}
\end{equation}
is the orbital bare density-density correlation function, and
 \begin{equation}
\chi^{0}_{n \mu}(q,\omega) = {1\over N}
\sum_{\sigma \sigma'} \langle n_{\sigma }(q) {\sum_{k}
c^{\dagger}_{k\sigma'}\Lambda^{\mu}(k,q)
c_{k+q\sigma' } }\rangle_{0}
\label{p1ab}
\end{equation}
A simple inspection of the
diagrammatic structure of the scattering amplitudes $\Gamma$
and of the response functions $P$ together with the 
observation that the static fermionic bubbles are non-singular
function of $q$, allows to conclude that both quantities provide the
same amount of information as far as the occurrence of 
the instabilities is concerned. In fact one can immediately recognize
that, since a diverging response function can only arise
 from a diverging
boson propagator, also the scattering amplitude mediated by the same
boson propagator would diverge at the same time. The same holds
for the static Cooper scattering amplitude in the particle-particle
channel, which, within our leading-order
$1/N$ expansion coincides with the particle-hole amplitude
\cite{notaextleg}.

\section{Physical properties of the Hubbard-Holstein model}

\subsection{Static properties}
The model introduced in the previous section is characterized by the
contemporary presence of a very strong local interaction and
a phonon-mediated attraction. The main point to be addressed
here is the subtle interplay between the attractive
and repulsive forces so as to clarify the origin of the effective
interactions arising between the quasiparticles
giving rise to instabilities for some values of the
parameters. 
We will first analyze the static properties, so that we first
focus on the $\omega=0$ limit of the effective interaction
in the particle-hole channel Eq.(\ref{gamma}) between
quasiparticles on the Fermi surface ($k=k_F$, $k'=k'_F$).

Since in general this quantity involves 
the calculation of the fermionic
bubbles (\ref{pi}) entering the expression of the boson propagator
(\ref{bosprop}) an explicit analytic evaluation of this scattering
amplitude at finite momentum is not possible. The finite momentum 
analysis of the static scattering amplitude
is then carried out numerically. However
we find instructive to present
first the analytic results, which can be obtained in the small
transferred momentum limit $(q\to 0$)
\begin{eqnarray}
\Gamma_q & = &  \lim_{q\to 0}\lim_{\omega \to 0} \Gamma 
(k_F, k'_F; q, \omega) \nonumber \\ 
 & = & -{1 \over {N}}
\left(
\begin{array} {c}
2E_F      \\
i         \\
-2g
\end{array} \right) \label{gam0} \\ 
& \times & \left(
\begin{array}{c c c}
\Pi^{rr}_q               & i2r_0^2 +\Pi^{r\lambda}_q & \Pi^{r a}_q    \\
i2r_0^2 +\Pi^{r\lambda}_q & \Pi^{\lambda\lambda}_q  &\Pi^{\lambda a}_q\\
 \Pi^{a r}_q             &\Pi^{a \lambda }_q
& 2\omega_0 + \Pi^{aa}_q
\end{array} \right)^{-1}
\left(
\begin{array} {c}
2E_F \\
i    \\
-2g
\end{array} \right) \nonumber
\end{eqnarray}
where the static $q\to 0$ limit of the bubbles,
$\Pi^{\mu \nu}_q \equiv  \Pi^{\mu \nu}(q\to 0, \omega_m=0)$
can easily be evaluated by noticing that at $T=0$
\begin{equation}
\Pi^{\mu \nu}(q\to 0,\omega_m=0)  = 
\sum_{k} {{\partial f\left( E_k \right)} \over
{ \partial E_k }}
{\Lambda}^\mu\left( k,0 \right)
{\Lambda}^\nu\left( k,0 \right) 
\nonumber
\end{equation}
\begin{eqnarray}
 & = & -\sum_{k} \delta\left(E_k-E_F \right)
{\Lambda}^\mu\left( k,0 \right)
{\Lambda}^\nu\left( k,0 \right) \nonumber \\
 & = & -\nu^* {\Lambda}^\mu\left( k_F,0 \right)
{\Lambda}^\nu\left( k_F,0 \right) \label{piq0}
\end{eqnarray}
where $\nu^*$ is the quasiparticle density of states at the
Fermi level.
In the last equality we use the property that the vertices
are constant over the Fermi surface. Once the expressions for
the bubbles are substituted in Eq.(\ref{gam0}) one obtains
the final expression for the static scattering amplitude 
between the quasiparticles
\begin{equation}
N\nu^*\Gamma_q={N\nu^*\Gamma_\omega \over {1+N\nu^* \Gamma_\omega}}=
{{4t\nu^*\varepsilon_{k_F} -\lambda_g} \over 
{1+\left(4t\nu^*\varepsilon_{k_F} -\lambda_g\right)}} \label{gamq}
\end{equation}
where we introduced the effective phonon-mediated $e$-$e$ coupling 
$\lambda_g\equiv 2\nu^*g^2/\omega_0$
In Eq.(\ref{gamq})
the dynamical effective scattering amplitude between the quasiparticles
$\Gamma_\omega=\Gamma (k_F,k_F;q=0,\omega \to 0)$ is identified by
\begin{equation}
\Gamma_\omega =
{1\over N}\left( 4t\varepsilon_{k_F} -{\lambda_g \over \nu^*}\right).
\label{gamdyn}
\end{equation}
This latter quantity represents the residual interaction between the 
quasiparticles on the Fermi surface when their mutual screening is not 
taken into account.
This screening effect is instead included in $\Gamma_q$ \cite{nozieres}.
The above expression for $\Gamma_\omega$ can also be easily obtained
from the direct evaluation of Eq.(\ref{gam0})
by noticing that in the dynamical limit the fermionic bubbles vanish
identically. Thus, using $\Pi^{\mu \nu}=0$ in Eq.(\ref{bosprop}) one
again finds  
$
N\nu^*\Gamma_\omega=4t\nu^*\varepsilon_{k_F} -\lambda_g.
$

At this point one recognizes that an instability can
in principle take place when the Landau-Pomeranchuk criterion
for the stability of a Fermi liquid $F_0^s \equiv N\nu^*\Gamma_\omega
>-1 $ is violated leading to a negatively diverging total (static)
scattering amplitude $\Gamma_q$
\begin{equation}
F_0^s=N\nu^*\Gamma_\omega=4t\nu^*\varepsilon_{k_F} -\lambda_g \le-1
\label{f0s}
\end{equation}
As a consequence a divergent compressibility is found
\begin{equation}
\kappa\equiv {\partial n \over \partial \mu}=
 \nu^* (1-\nu^* \Gamma_q)={N\nu^* \over 1+F_0^s }{\to_{F_0^s\to -1} \infty}
. \label{compressibility}
\end{equation}
which signals the occurrence of a PS.
It is worth noting that the phonon parameters only enter the
condition for PS via the combination $\lambda_g=2\nu^*g^2/\omega_0$.
This implies that, for a given electronic band structure,
i.e. for a given $\nu^*$, the {\it static} instability
 is obtained for any phonon 
frequency provided $g$ is suitably rescaled to keep the
$\lambda_g$ fixed (of course too large $e$-$ph$ couplings
put in jeopardy our weak coupling approach, where vertex
correction beyond the Migdal theorem are not included).

 It is worth remarking that the quantity $\varepsilon_{k_F}=
 \left(\cos k_{xF} +\cos k_{yF} \right)
+2\alpha \left(\cos k_{xF} \cos k_{yF} \right)$ is rather small at
low doping \cite{ef}.
We carried out a detailed investigation of $\varepsilon_{k_F}$
as a function of the doping and of the hopping ratio $\alpha$
determining the most appropriate values of $t'$
in order to reproduce within our single-band model Fermi surface 
shapes in reasonable agreement with those observed in 
some superconducting copper oxides. 
According to previous analyses \cite{fukuyama},
we found that the values of $t'$ giving rise
to reasonably shaped Fermi surfaces for 214 compounds are negative with 
$t'\simeq -0.167t $. $t'\simeq -0.45 t$ was instead adopted to reproduce 
Fermi surfaces in agreement with those observed in 123 and 2212
compounds. For these values of $t'$, $\varepsilon_{k_F}$
remains rather small in the low doping regime. Therefore, although
a minumum critical value $g_c$ of the $e$-$ph$ coupling is 
needed for the phase-separation instability to occur, its value
can be small and it does not really provide a difficult condition
to be satisfied in the real systems \cite{notaefgt0}. 

The main point to be stressed here is that, despite the infinite
bare repulsion between the bare particles, a huge screening takes
place in the system introducing attractive forces almost exactly
balancing the repulsion and giving rise to a finite scattering 
amplitude $\Gamma_\omega$ between the quasiparticles. 
This residual effective interaction is
repulsive in the absence of the $e$-$ph$ coupling, whereas it may
turn into an attraction when $\lambda_g$ is large enough. 
In this situation, which arises from the strongly interacting
nature of the system, even this additional
 attraction can drive the system unstable. This finding
matches well previous results obtained in different
models and it emphasizes the general robustness of the PS
concept in the context of strongly interacting systems.
Moreover this finding supports the choice of the Hubbard-Holstein
model as a simple paradigm to describe the physics of the
PS in the context of strongly interacting systems.

In principle the above analysis cannot exclude that
an instability  at a finite momentum can take place
before PS, thus calling
for a more general  investigation. This latter has
been carried out numerically through the explicit calculation
of both the static scattering amplitude at finite transferred
momentum and the density-density response function. 
In Fig. 2 we report the full static scattering amplitude
in the particle-particle channel.
\begin{figure}
{{\psfig{figure=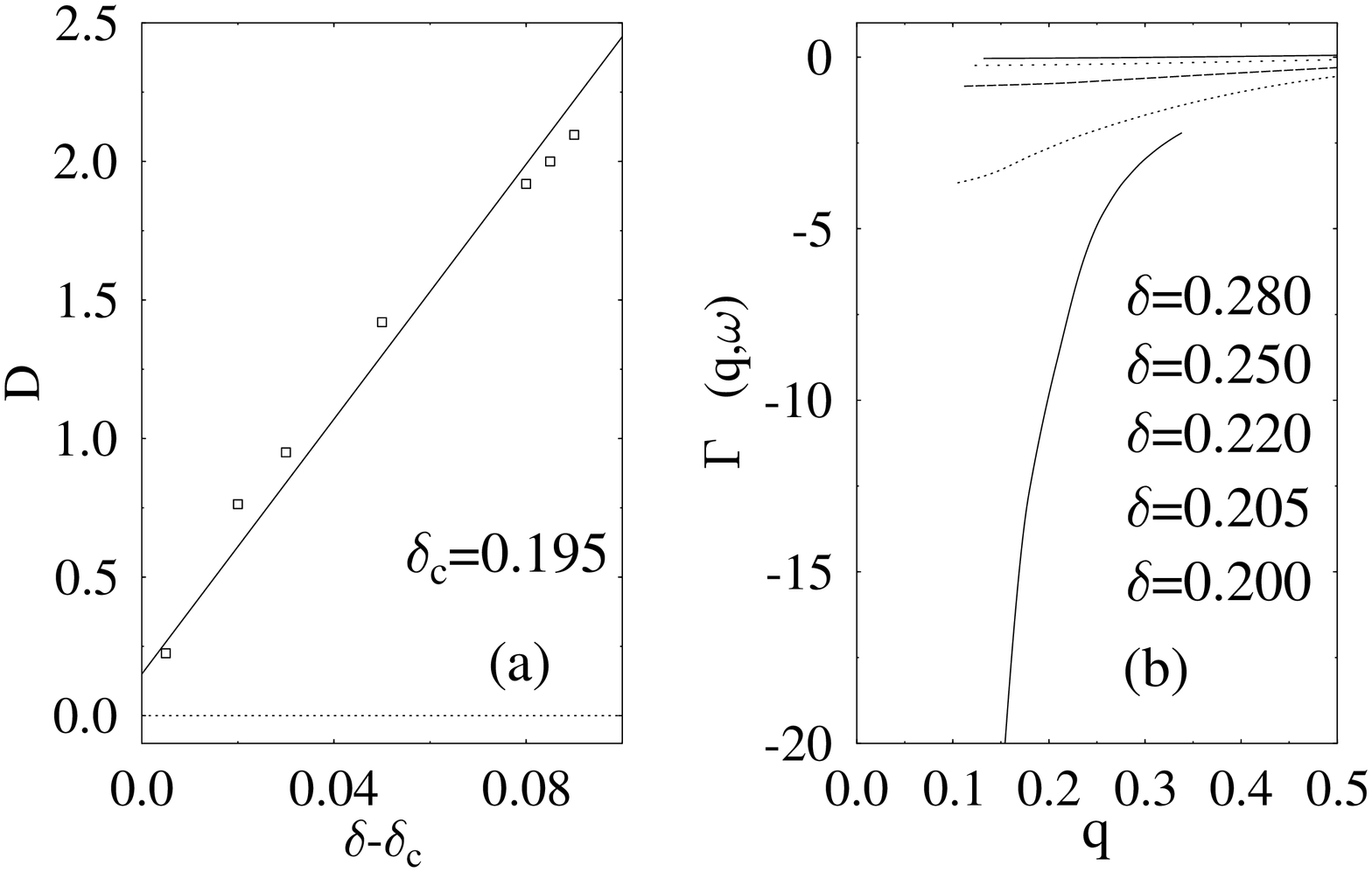,width=5.5cm,angle=-90}}}
{\small Fig 2. (a) Doping dependence of the mass $D$ in the static
effective scattering amplitude close to the PS instability.
The open squares in (a) indicate the values of $D$ at various dopings
and $g_{\rm {phys}}=g_{c \,{\rm phys}}=0.194/\sqrt{2}$ eV, 
$t_{\rm phys}=0.5$eV, $t'=-(1/6)t$, and 
$\omega_{0\, {\rm phys}}=0.04$eV. For these parameters $\delta_c=0.195$.
The solid line is a linear fit. (b) Static scattering amplitude
for the same parameters as in (a) as a function of the
transferred momentum $\mbox{\boldmath $q$}$ in the (1,0) 
direction. The doping $\delta=0.2,0.205, 0.22,0.25, 0.28$ increases
from the lower solid line to the upper solid line.}
\end{figure}
We choose this quantity because
it will also enter the calculation of the Cooper instability
reported below. We performed the calculation for various
different values of the doping at a value of $g$ larger than
the minimum value required to have a PS instability.
Here and throughout this paper we express the various
quantities in physical units translated from the
$1/N$ formalism with $N=2$: 
$t_{\rm phys}=t/N$, $t_{\rm phys}'=t'/N$, 
$g_{\rm phys}=g/\sqrt{N}$, and $\omega_{0\rm phys}=\omega_0$.

From the reported results it is natural to conclude that the divergent
particle-particle 
scattering amplitude (i.e. a divergent boson 
propagator and a consequently
diverging density-density response function) occurs at a zero momentum
transfer. According to Eq.(\ref{compressibility}) this divergency
($\Gamma_q \to -\infty$) leads to a diverging compressibility
signaling a PS instability.

As it can be seen in the 
$g$ vs $\delta$ phase diagram of Fig. 3,
for the $t'=-0.167 t$ case (a similar diagram is obtained
when $t'=-0.45 t$), we found that this behavior is rather generic 
at low and intermediate doping (continuous line)\cite{notacdwsr}. 
\begin{figure}
{{\psfig{figure=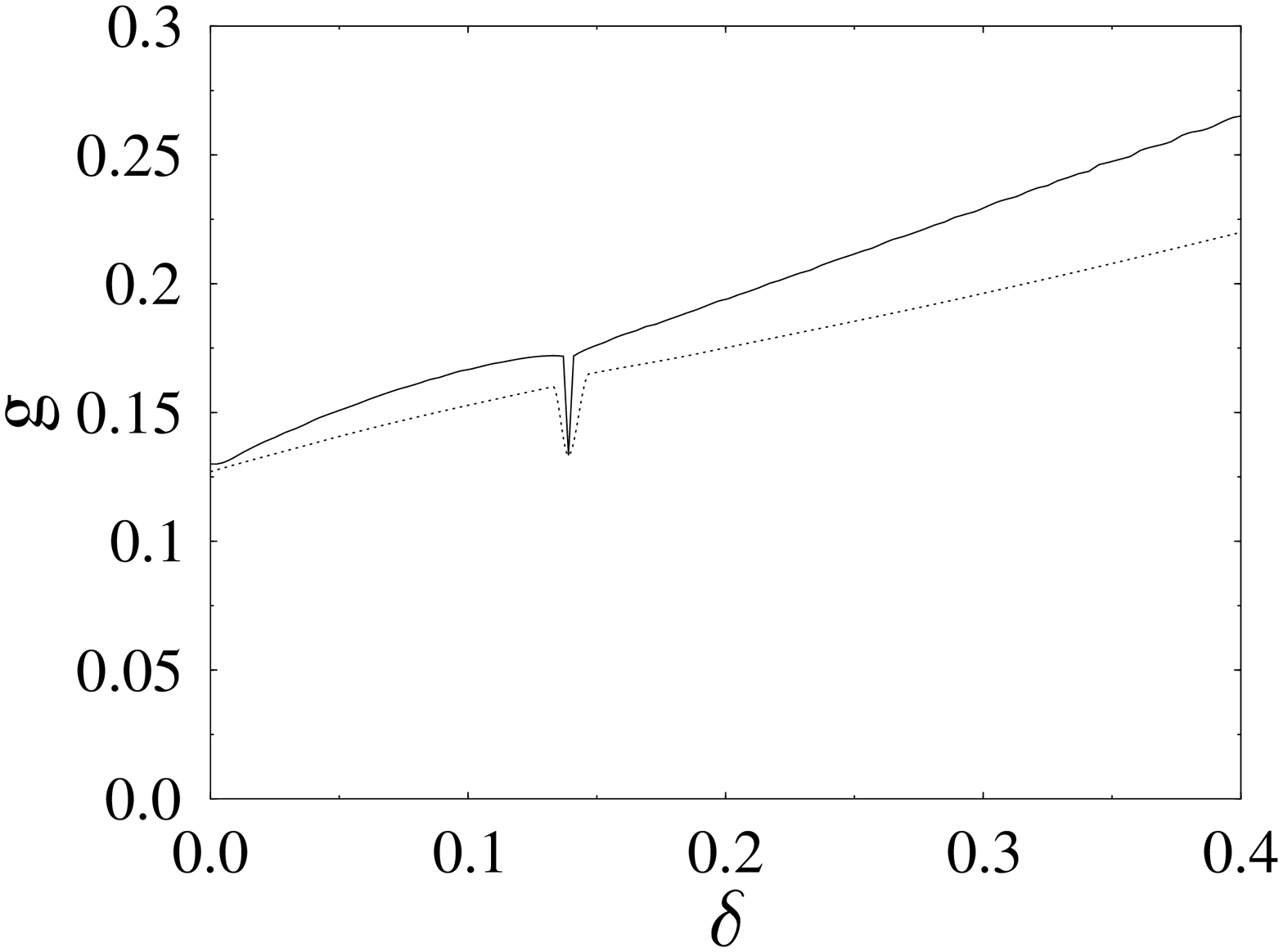,width=5.5cm,angle=-90}}}
{\small Fig 3. Phase diagram $e$-$ph$ coupling $g$ vs doping $\delta$
$t_{\rm phys}=0.5$eV, $t'=-(1/6)t$, and 
$\omega_{0\, \rm phys}=0.04$eV. On the solid line the compressibility
diverges, whereas the dotted line arises from the Maxwell construction.}
\end{figure}
The instability line drawn in the phase diagram 
indicates where the static density-density response of the system
becomes singular. A narrow dip in the instability solid line 
is due to the presence of a van Hove singularity enhancing
$\nu^*$, thus favoring the instability.
However, the system becomes unstable and phase
separates before the solid line is reached,
where $\kappa \to \infty$. A Maxwell construction (dashed line)
is needed to determine the region where PS starts.
Appendix A briefly describes the procedure to carry out the
Maxwell construction.

We complete the static analysis of the model by
investigating the possibility of Cooper pairing.
As already pointed out, in various models with strong
$e$-$e$ correlations superconductivity can appear in the proximity
of a PS instability. This can easily be
understood by looking at the large attraction arising
in the particle-particle effective scattering amplitude
close to the instabilities (see Fig.2).
According to this
simple observation and according to the previous experience in
other strongly interacting models, we therefore investigated the
Fermi surface average of the particle-particle scattering amplitude
defined in Eq.(\ref{cooper}) 
\begin{eqnarray}
\lambda_l &=& - \left[\int dk \delta \left( E_k-\mu \right) g_l(k)^2 
\right]^{-1} \nonumber \\
& \times & \int \int dk dk' \left[ 
g_l(k) \Gamma^C \left( k,k';\omega=0 \right) g_l(k')
\right. \nonumber \\
& \times & \left. \delta \left( E_k-\mu \right)
\delta \left( E_{k'}-\mu \right) \right] 
\label{coupl}
\end{eqnarray}
with $g_{s_1}(k)=\cos(k_x)+\cos(k_y)$, and
$g_{d_1}(k)=\cos(k_x)-\cos(k_y)$ 
projecting the interaction onto the $s$-wave and $d$-wave channels.
(Notice that $\lambda_l>0$ means attraction).

The results are tabulated at various doping 
concentrations for the case with 
$t_{\rm phys}=0.5$eV, $t=-1/6t'$, $g_{\rm phys}=0.192/\sqrt{2}$eV,
$\omega_{0\rm phys}=0.04$eV in Table I. 
\begin{table} 
\begin{center}
  \begin{tabular}{|c|c|c|c|c|c|}
   \hline
  $\delta$        & 0.194 & 0.200  & 0.205 & 0.210 & 0.225 \\
  $\lambda_{s_1}$ & 0.458 & -0.065 &       &       &       \\
  $\lambda_{d_1}$ & 1.284 & 0.137  & 0.092 & 0.076 & 0.052 \\
   \hline
   \end{tabular}
  \end{center}
{\small Extended $s$- and $d$-wave superconducting couplings 
for $t_{\rm phys}=0.5$eV, $t=-1/6t'$, $g_{\rm phys}=0.192/\sqrt{2}$eV,
$\omega_{0\rm phys}=0.04$eV at various dopings. The instability
is at $\delta_c=0.193$.}
\end{table}   
With the set of parameters of Table I
the critical doping for the occurrence of the instability 
is $\delta_c=0.192$. 
Whereas the couplings $\lambda_d$ are found to be generally
attractive near (and inside) the unstable region,
$s$-wave Cooper instabilities are found only very close to the
instability line\cite{notacdwsrb}.  

At first sight it may seem strange to investigate and to look for
superconductivity in a phase-diagram region, which is made
unaccessible by the Maxwell construction. However, one should
remember that the phase diagram of Fig. 3 may be modified by 
several effects. First of all long-range Coulombic forces
will spoil PS. This issue is the main topic of the present
paper and will be discussed in Section IV. Secondly
temperature effects could restore the uniformity of the
system. This issue cannot be reliably addressed within
a slave-boson formalism and is beyond the scope of our
work. Most importantly superconductivity itself
can give rise to a more stable phase competing with
and stabilizing PS. In this latter scenario a complex
interplay between the PS instability and superconductivity
will likely arise: The incipient instability originates
superconductivity, which, in turn, prevents the
instability to occur\cite{notacccdgr}. To substantiate
these ideas a consistent analysis would be required 
of the feedback effects, which, however, only appear
at higher order in the 1/N expansion. Work in this
direction is in progress.

\subsection{Dynamical properties}
The analysis carried out in the previous sections was purely
static. From this analysis it turned out that the subtle
balancing between repusive forces and attractive screening processes
may have relevant physical effects. In this regard it immediately
appears of great interest a dynamical investigation in order to fully
clarify the way the various screening effects contribute to the
physics of the system. 

From the standard Fermi-liquid theory \cite{nozieres}, 
one obtains the two relations connecting the density vertex  
$\Lambda^e(q,\omega)$ and the wavefunction renormalization 
$z^e$ in the dynamic and static limits 
\begin{eqnarray} 
z^e\Lambda^e\left(q=0,\omega \to 0\right) & = & 1  \label{wi1}\\ 
z^e\Lambda^e\left(q\to 0,\omega = 0\right) & = &
{{1}\over{1+F_0^{s(e)}}}\,\, . \label{wi2}
\end{eqnarray}
where $F_0^{s(e)}=2\nu^*\Gamma^e_\omega$  
and $\Gamma^e_\omega$ are the Landau parameter
and the
dynamic ($q=0$, $\omega \to 0$) effective $e$-$e$ scattering amplitude
between the quasiparticles respectively, when only the $e$-$e$ repulsion is
taken into account. To explicitly keep
memory of this limitation we append a suffix ``$e$" to
$\Gamma_\omega$ and to any quantity not involving phononic processes.

In our specific model and within our leading-order $1/N$ approximation
one can recognize that the static and dynamical
$e$-$ph$ vertices, $\Lambda_q$ and $\Lambda_\omega$,
are related by 
\begin{equation}
\Lambda_q^e={\Lambda_\omega^e
\over 1+N\nu^*\Gamma^e_\omega}\,\, ,
\label{vst}
\end{equation}
where $\Gamma^e_\omega$ is the first term in the r.h.s. of
Eq.(\ref{gamdyn}) and $\Lambda_\omega^e=1$.
Eq.(\ref{vst}) shows the difference
between the dynamic and static limits,
which can be substantial for large $\nu^*$. 
This is a generic feature of strongly correlated systems,
which one has to take into account since it can strongly affect
the relevance of the $e$-$ph$ coupling. 

The above small-$q$ and small-$\omega$ analysis can easily be 
extended to finite momenta and frequencies by an explicit numerical
evaluation of the diagrams of Fig. 4. 
\begin{figure}{{\psfig{figure=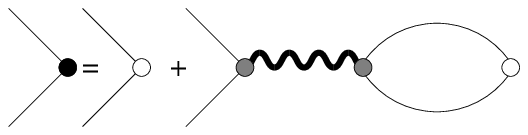,width=3.5cm,angle=-90}}}
{\small Fig 4. Leading-order in $1/N$ diagrammatic structure of the
effective $e$-$ph$ vertex dressed by electronic processes only:
the wavy line is the slave-boson propagator only involving
$r$ and $\lambda$ bosons, the solid dot is the dressed $e$-$ph$
vertex, the open dot is the bare $e$-$ph$ vertex and
the grey dots are the quasiparticle-slave-boson vertices. }
\end{figure}
In these diagrams one
considers the effects of the $\delta r_i$ and $\delta \lambda_i$
boson fluctuations, keeping track at leading order in $1/N$
of the original infinite Hubbard repulsion U. More explicitly one
has to evaluate 
\begin{eqnarray}
 \Lambda^{\rm e-ph}(k,q;\omega_m)& = & \nonumber \\
 \Lambda^a(k,q) - N\sum_{\mu,\nu =r,\lambda}
& \Lambda^\mu(k,q) &  D^{\mu,\nu}(q,\omega_m)\Pi^{\nu,a}(q,\omega_m)
\label{vqw}
\end{eqnarray}
For $\omega_m=0$ our slave-boson result is in perfect 
quantitative agreement with the
static analysis carried out in 
Refs.\cite{zeyher,zeyher2} for the single-band
infinite-U  Hubbard model treated within a large-$N$ 
expansion by means of Hubbard projectors. 

The results of a dynamical analysis of the
$e$-$ph$ vertex vs transferred Matsubara frequencies
are reported in Figs. 5(a)
and 5(b) for $t'=-t/6$ for a small [${\mbox{\boldmath $q$}}=(0.2,0)$], 
and a sizable [${\mbox{\boldmath $q$}}=(2.0,0)$] value of the transferred
momentum respectively. Similar results are obtained for the $t'=-0.45t$
case. It is worth noting in Fig. 5(a) how rapidly the effective
$e$-$ph$ vertex increases as soon as the transferred frequency
becomes larger than some screening scale $\omega_{\rm scr}$
of the order of $v_F^* q$, with $v_F^* \propto \delta$.  
A much larger scale of the order of
the bare bandwidth $t$ is involved in the 
slower increase of the vertex at large momenta [Fig. 5(b)].
A detailed discussion of the screening scales is deferred to
the next section, where this analysis will also be carried out in the
presence of LRC forces.
\begin{figure}
{{\psfig{figure=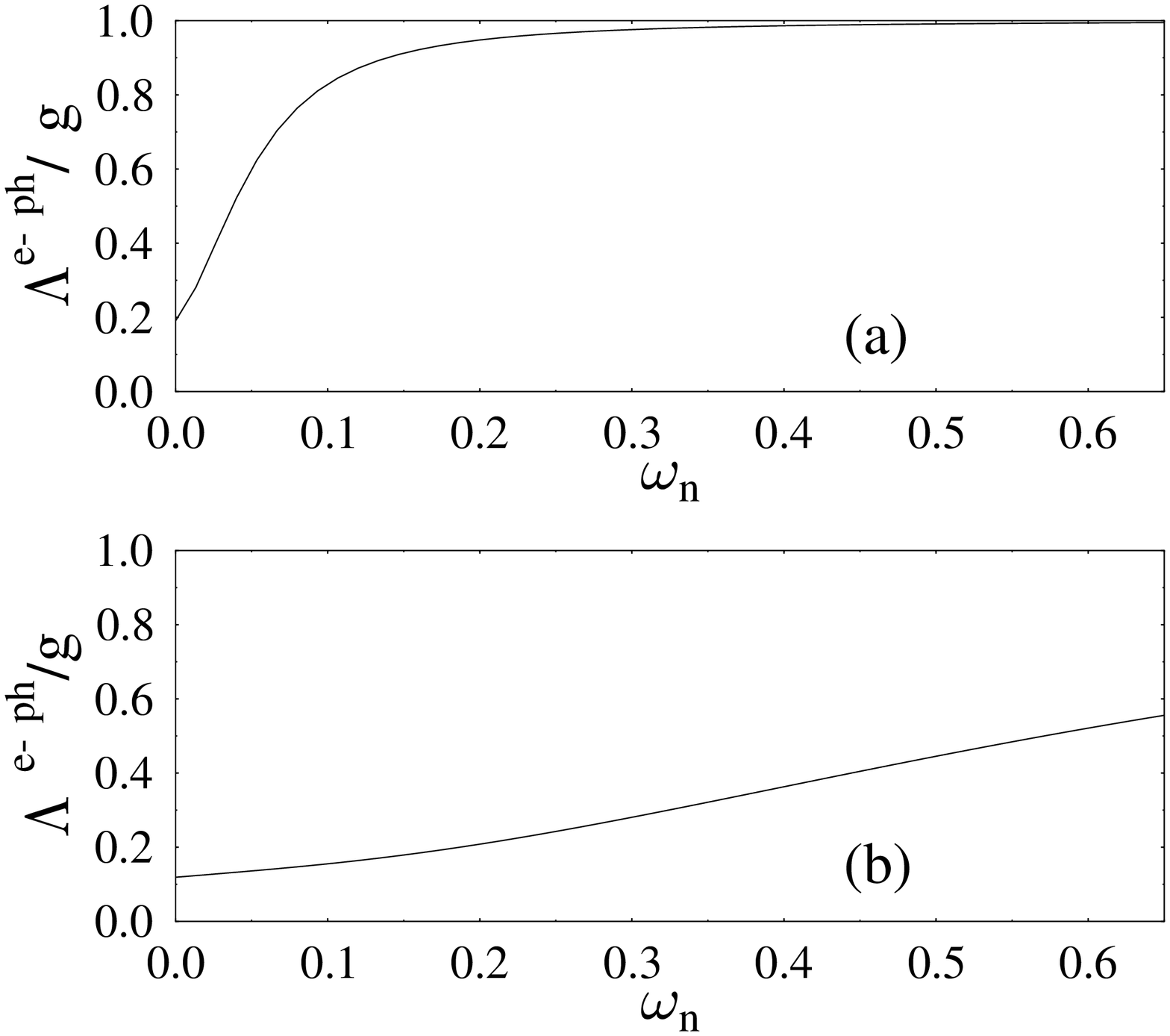,width=5.5cm,angle=-90}}}
{\small Fig 5. Effective $e$-$ph$ vertex as a function of the transferred
Matsubara frequency for $t_{\rm phys}=0.5$eV, $t'=-(1/6)t$, and 
$\omega_{0\, \rm phys}=0.04$eV. In (a) the transferred momentum
is small, ${\mbox{\boldmath $q$}}=(0.2,0)$, and it is large,
${\mbox{\boldmath $q$}}=(2.0,0)$, in (b)}
\end{figure}
\begin{figure}
{{\psfig{figure=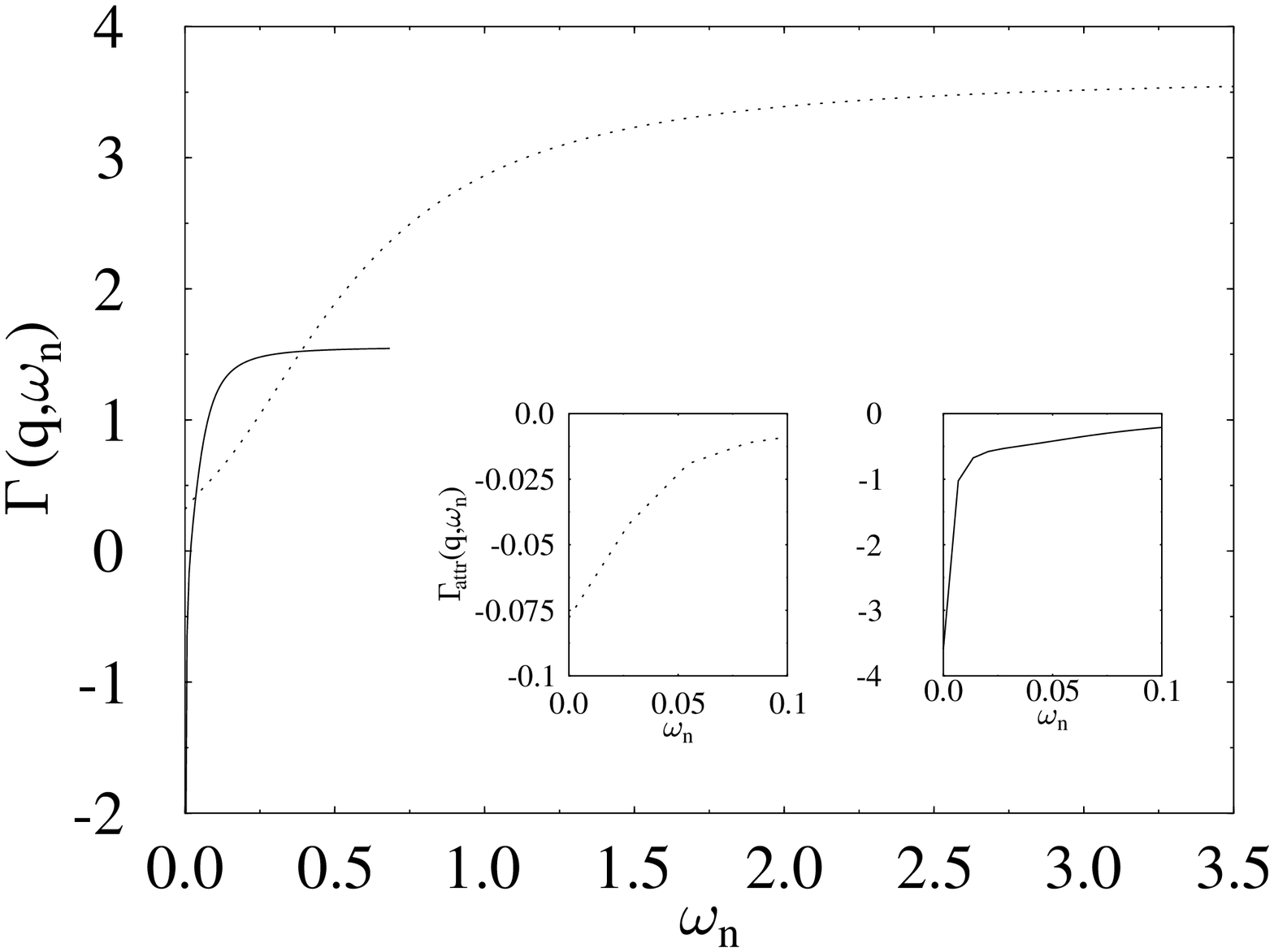,width=5.5cm,angle=-90}}}
{\small Fig 6. Effective scattering amplitude as a function of 
the transferred Matsubara frequency for 
$t_{\rm phys}=0.5$eV, $t'=-(1/6)t$, $g_{\rm phys}=
0.205/\sqrt{2}$eV, and $\omega_{0\, \rm phys}=0.04$eV.
Solid line: Small transferred  momentum ${\mbox{\boldmath $q$}}=(0.15,0)$;
dashed line: Large transferred momentum 
${\mbox{\boldmath $q$}}=(1.75,0)$. In the insets the attractive
part $\Gamma_{\rm attr}=\Gamma-\Gamma_{\rm rep}$ (see text)
are reported for both small (solid line) and large momenta (dashed line).}
\end{figure}
Since dynamical effects strongly modify the behavior of the 
$e$-$ph$ interaction, 
it seems natural to extend the dynamical analysis
to other relevant quatities of the system.
In particular we investigated the finite frequency behavior 
of the effective Cooper scattering amplitudes between the quasiparticles
on the Fermi surface $\Gamma^C(k_F,k_F',\omega_n)$.
The results are reported in Fig. 6 for the particle-particle
scattering amplitudes as a function of the transferred Matsubara
frequencies both for small and large momenta.
In this case the bare $e$-$ph$ coupling $g_{\rm phys}=0.194/\sqrt{2}$eV 
and the doping $\delta=0.205$ are tuned in order to place 
the system in the proximity of a $q=0$ instability occurring
for $g_{c \,\rm phys}=0.194/\sqrt{2}$ and $\delta_c=0.195$.
For clarity we also represent in the inset 
the difference between the total and the 
purely repulsive (i.e. involving
the $r$ and $\lambda$ bosons only) part of the scattering 
amplitude. For any momentum, this attractive part of the interaction
lives on a frequency range of the order of the phononic energy.
However, as expected, the large and the small momentum 
behavior are strikingly different.
At small momenta the attraction mediated by the
phonons is stronger because the $e$-$ph$ coupling is larger
(cf. Fig. 5) and, close to the instability, it gives rise
to a large attraction at low frequencies.
This attraction is the vestige of the huge static attraction
found close to the PS instability also in the particle-particle
channel (cf. Fig.2). On the other hand the attractive
contribution in the large-momentum case (dotted curve in the
inset of Fig. 6) is quit small for any frequency, because
the effective coupling between the quasiparticles and the
phonons is greatly reduced by the large-momentum screening
[cf. Fig.5(b)].

We finally like to comment on the behavior of the scattering
amplitude for frequencies that are larger than both the phononic and
electronic energy scales. In this case it can be easily checked
that the fermionic polarization bubbles vanish as $\omega_n^{-2}$
thereby leading to a scattering amplitude determined by the
bare slave-boson propagators [the bare phonon propagator 
$(B^{a,a})^{-1}$ also vanishes for large frequencies].
A simple calculation based on Eq.(\ref{cooper}) and on Eq.(\ref{bosprop})
with $\Pi(q,\omega)\to 0$ gives the large-$\omega$ saturation value
of the scattering amplitude in the Cooper channel
\begin{equation}
\Gamma^C(k_F,k_F';\omega \to \infty)={1\over N}\left[
4t\varepsilon_{k_F}+{\lambda_0^0\beta_q +\alpha \lambda_0^1 \gamma_q
\over 2r_0^2} 
\right]
\end{equation}
where $\lambda_0^{0,1}$ are defined in Eq.(\ref{lambda}) and
$\beta_q$ and $\gamma_q$ are given after Eq.(\ref{lambda}).
A direct inspection of Fig. 6 shows that this saturation value
is reached more rapidly in the small momentum transfer case,
whereas a slower rise of $\Gamma^C(\omega)$ is found
in the large momentum transfer case of Fig. 6. This result
is a natural consequence of the disappearance of the phonons from
the high frequency processes at ($\omega_n > \omega_0$). Then
only electronic processes determine the scattering like in the screening
of the $e$-$ph$ vertex described above [cf. Figs 5(a),(b)]. 
Again two energy scales appear to be 
relevant: A small $\omega_{scr}\sim 
\delta q$ at small momenta and a large $\omega_{scr}\sim t$ 
at large momenta.
These two scales set the frequency region above which the scattering
tends to saturate. 

As far as the dynamic behavior of the scattering amplitude
in the proximity of a PS instability, we also like,
for completeness sake, to remind that
a recent work \cite{CDGPRL} also showed that, close to a $q=0$
instability, the effective dynamical scattering amplitude
for real frequencies has a strongly singular behavior.
In particular the effective interaction assumes the same form
as the one obtained within the gauge-theory treatment
of the t-J model \cite{nagaosa}
\begin{equation}
\Gamma (q,\omega) \approx \tilde{U}
- {1 \over Bq^2 - i\omega {C\over q} +D} \label{fitgamsr}
\end{equation}
with a mass $D\propto (\delta-\delta_c)$, which vanishes linearly when,
by varying the doping one approaches the $q=0$ instability line.
$\tilde{U}$ represents the almost momentum independent
repulsive contribution to $\Gamma$ mediated by the 
$r$ and $\lambda$ bosons. According to the
spirit of the Fermi liquid theory, it may be interpreted
as the residual repulsion surviving between the
quasiparticles and arising from the infinite repulsion U between
the bare electrons. Eq.(\ref{fitgamsr})establishes a connection
between the presence of a $q=0$ charge instability and singular
scattering, which could determine
the anomalous normal state properties of the copper oxides.

The analysis summarized in Figs. 5 and 6 is of obvious pertinence 
in a complete Eliashberg treatment of the superconductivity
problem. In particular it is evident that the huge enhancement 
of the attractive part of the scattering amplitude 
near the instability line can be
responsible for large critical temperatures despite the small
e-ph coupling. 

The closeness to a phase-separation instability appears therefore as 
a favorable condition in order to obtain high temperature
superconductivity from a phonon-mediated attraction 
similarly to what suggested in the context of purely 
electronic pairing mechanisms\cite{GRCDK1,GC}.

\section{The effect of long-range interactions}
\subsection{Formal extension}
Although we expect the effect of LRC forces to be most effective
in the small transferred momentum case, where the underlying lattice
structure is less visible, we  explicitely
kept into account the real symmetry of the square-lattice system. 
Therefore, to derive an explicit expression for the Coulombic potential
in the spirit of the point-charge approximation
we started from the discretized form of the Laplace equation.
Moreover, since we assume that our twodimensional model
represents planes of a truely threedimensional lattice
we also include a third spacial dimension. For clarity
in this section we restore the explicit dependence of the 
lattice spacing $a$, which in the previous sections was
set to unity in the square twodimensional lattice. In the 
third space direction, instead, we assume the unit cell to
have a lattice spacing $d$ (i.e. we assume a tetragonal
threedimensional lattice). In this scheme, the Laplace
equation reads
\begin{equation}
\epsilon_{\parallel}\sum_{\eta=x,y} 
\left( {\phi_{i-j+\eta}+ \phi_{i-j-\eta}-2 \phi_{i-j}
\over a^2 } \right) \nonumber
\end{equation}
\begin{equation}
+
\epsilon_{\perp}
\left( {\phi_{i-j+z}+ \phi_{i-j-z}-2 \phi_{i-j}
\over d^2 } \right)=
-e \delta (i-j)
\end{equation}
where $i,j$ are lattice sites and $\epsilon_{\perp}$ and
$\epsilon_{\parallel}$ are the dynamic dielectric constants
perpendicularly  and along the planes respectively.
Fourier transforming one can easily obtain
\begin{eqnarray}
\left[ {\tilde{\epsilon}\over (a/d)^2 } \left[ 
\cos (aq_x) + \cos (aq_y) -2 \right] + \cos (dq_z) -1 
\right] \nonumber \\
\times {2\epsilon_{\perp}\over d^2} 
\phi_{\bf q} 
= -e 
\end{eqnarray}
with $\tilde{\epsilon}\equiv \epsilon_{\parallel}/
\epsilon_{\perp}$, from which one gets the expression of
the LRC potential in the threedimensional momentum space
\begin{equation}
\phi_{\bf q} = -{e d^2 \over 2 \epsilon_{\perp}}
\left[ A(q_x,q_y) + \cos (q_zd) \right]^{-1}
\end{equation}
where we defined
\begin{equation}
A(q_x,q_y) = {\tilde{\epsilon}\over (a/d)^2 } \left[ 
\cos (aq_x) + \cos (aq_y) -2 \right] -1\,\,.
\end{equation}
Since we are interested in the effects of the Coulomb potential
on the square lattice planar system, we now transform from
$q_z$ to the real space for the plane at $z=0$ obtaining
\begin{equation}
\phi_{{\bf q}_\parallel} (z=0)= -{e d \over 2 \epsilon_{\perp}}
{1 \over \sqrt{A^2(q_x,q_y) -1}}\,\,.
\end{equation}
Notice that this is the potential between electrons 
in a twodimensional
lattice embedded in a threedimensional space and it diverges
as $q^{-1}$ for small transferred momenta, rather than $q^{-2}$ 
as it happens for threedimensional electronic systems.
This potential can be used in the Coulombic part of the Hamiltonian
\begin{equation}
H_C = {V_C \over 2 N} \sum_q {1 \over \sqrt{A^2(q) -1}}
\rho_q \rho_{-q} \label{hamc}
\end{equation}
where $\rho_q \equiv \sum_{k,\sigma} c^\dagger_{k+q,\sigma}c_{k,\sigma}$
and the Coulombic coupling constant $V_C\equiv 
e^2 d /(2 \epsilon_{\perp} a^2 )$.
It should be noted that, as it is customarily done, 
the sum does not include the zero-momentum component, since
we are supposing that the diverging $q=0$ interaction between the
electrons is canceled by the contribution of a uniform positively
charged ionic background.
Having in mind the superconducting copper oxides
of the 214 type, where $d \approx 3 a$,
$t_{\rm phys}\approx 0.5$eV, $\epsilon_{\parallel}\approx 30$
 and $\epsilon_{\perp}\approx 5$ one sees that
$V_C$ has to range from roughly 0.5eV to 3eV in order to have 
holes in neighboring ${\mathrm {CuO_2}}$ cells repelling
each other with a strength of 0.1-0.6eV.

The Hamiltonian (\ref{hamc}) can then be added to the Hamiltonian
of the Hubbard-Holstein model (\ref{ham2}) and the product
of four fermionic fields can be decoupled by means of a standard
Hubbard-Stratonovich transformation. In this way one introduces
a new real bosonic field $Y_i$ to be integrated over in the functional
integral. Although this spacially 
fluctuating field does not have its own dynamics
it acquires a frequency dependence via its coupling to the
fermionic degrees of freedom. In particular, extending the
bosonic space $A^\mu(q)=( \delta r_q, \delta \lambda_q, 
a_q, Y_q)$ (we dropped for simplicity the Matsubara frequency dependence
of the bosonic fields) one can extend the formalism 
of Section II to include the effects of the Coulomb potential
represented by the $Y$ boson. A direct calculation shows
that the quasiparticle-boson vertex is
\begin{equation}
\Lambda^Y(k,q)=i
\end{equation}
as expected since the Coulomb potential couples to the local electronic 
density in the same way as the boson $\lambda$ does.
The bare boson propagator becomes a $4\times 4$ matrix
with an additional non-zero element: 
\begin{equation}
B^{Y,Y}=
{\sqrt{A^2(q) -1}  \over 2 V_C}
\end{equation}
We stress again that the $Y$ boson
does not have any uniform $q=0$ component, which was discarded
from the beginning in the sum of Eq.(\ref{hamc}) and therefore
it does not affect the mean-field results.

\subsection{Static properties}
The introduction of a LRC potential greatly affects the
coupling between phonons and electrons particularly
for small momenta, where the Coulombic forces are most
effective. In fact, since in the Hubbard-Holstein model phonons
couple to the local electron densities like the Coulombic
potential, the $e$-$ph$ coupling is effectively screened by 
particle-hole pairs created by the $e$-$e$ Coulombic
scattering processes and the effective 
$e$-$ph$ vertex is largely suppressed in the small-momentum
limit. This effect appears in a direct
calculation of the expression (\ref{vqw}) for the
effective $e$-$ph$ vertex
in the presence of LRC forces,
$ \Lambda^{\rm e-ph}_{LR}(k,q;\omega_m)$.
Now $\mu,\nu =r,\lambda ,Y$  in  (\ref{vqw}), so as to
include the screening due to the fluctuations of the
``long-range'' boson $Y$. In particular Fig. 7 shows
the momentum dependence of the static ($\omega_n=0$)
$e$-$ph$ vertex. The comparison with the same quantity
in the absence of LRC forces shows a striking difference
in the low-momentum region, where the Coulombic screening 
leads to a vanishing $e$-$ph$ vertex.
\begin{figure}
{{\psfig{figure=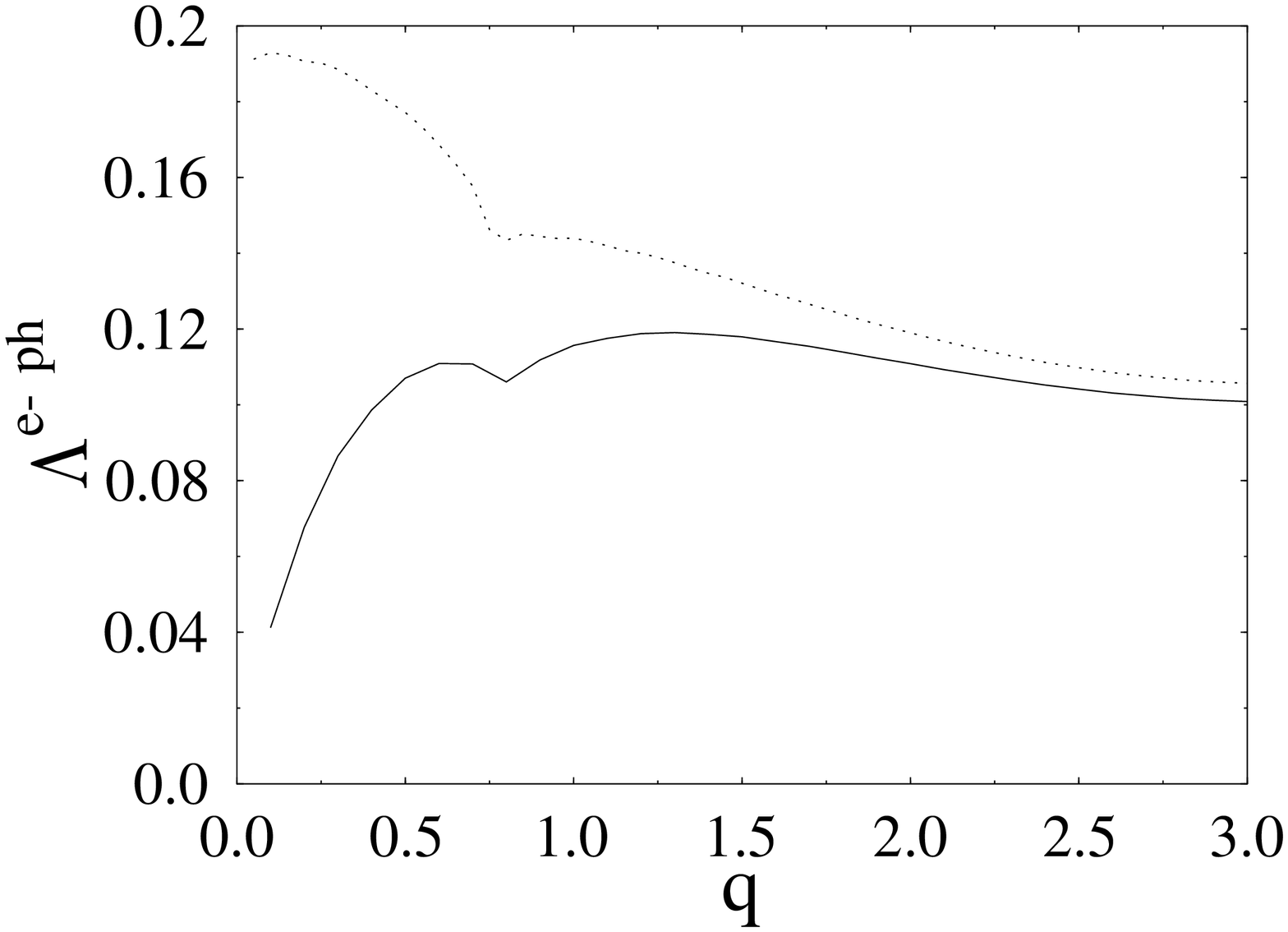,width=5.5cm,angle=-90}}}
{\small Fig 7. Static effective $e$-$ph$ vertex as a function of 
the transferred momentum in the (1,0) direction for
$t_{\rm phys}=0.5$eV, $t'=-(1/6)t$,  
$\omega_{0\, \rm phys}=0.04$eV. The dotted line is in the
absence of LRC forces ($V_{C\rm phys}=0$); the solid line
is in the presence of LRC forces with $V_{C \rm phys}=0.55$eV}
\end{figure}
This difference can easily
be understood by a three-step evaluation of
$ \Lambda^{\rm e-ph}_{LR}(k,q;\omega_m)$.
One can  first go through
the calculation of the vertex in the absence of LRC forces
[Eq.(\ref{vqw})]. Then one can introduce a density-density
fermionic bubble dressed by all electronic short-range
processes
\begin{equation}
\Pi^{eSR}(q,\omega_m)={4r_0^2P^0(q,\omega_m) \over 
\det {D^{2\times 2}}(q, \omega_m) }
\end{equation}
where $D^{2 \times 2}$ is the $2 \times 2$ sector of the
boson propagator only including the $r$ and $\lambda$
bosons, representing the purely electronic short-range
processes. Finally one performs a resummation
only including $Y$-boson fluctuations dressed by the $\Pi^{eSR}$
density-density fermionic bubbles.
The result is 
\begin{equation}
 \Lambda^{\rm e-ph}_{LR}(k,q;\omega_m)
={\Lambda^{\rm e-ph}(k,q;\omega_m)
\over 1+
\Pi^{eSR}(q;\omega_m)
{V_C \over \sqrt{A^2(q) -1}}}
\,\,. \label{ephlr}
\end{equation}
This expression readily shows the suppression of the
short-range-only $e$-$ph$ vertex appearing in the numerator
and clearly displays the vanishing of the LR vertex
for $q \to 0$, when the Coulombic potential diverges
$(A^2(q) -1)^{-1} \to \vert {\bf q} \vert^{-1} $.

The important physical consequence of the above suppression of the
effective $e$-$ph$ vertex is that in the presence of LRC forces
the long-wavelength density fluctuations are decoupled from the
phonons and become unable to drive a low-momentum instability.
Nevertheless the possibility of finite-momentum instabilities
still remains open and a detailed reanalysis of the phase 
diagram in the presence of the LRC potential is needed.
This analysis can be carried out by direct numerical
evaluation of the {\it static}
density-density response function and of its
divergencies. A direct inspection to the diagrammatic structure of the 
leading-order in $1/N$ 
density-density response function shows that one can include
the Coulombic effects both by extending the $\mu,\nu$ summations 
in Eq. (\ref{pab}) to include the $Y$ boson or, alternatively,
by first calculating the short-range response function
$P^{SR}$ by only including the $r,\lambda$ and $a$ bosons
and then to resum with the bare $\langle YY \rangle$ propagator.
The result is then
\begin{equation}
P^{LR}(q,\omega=0)={P^{SR}(q,\omega=0) \over 1+
{V_C \over N \sqrt{A^2(q) -1}}P^{SR}(q,\omega=0) 
   }
\,\,. \label{ddlr}
\end{equation}
From this expression one can easily see that a positively diverging
$P^{SR}(q,\omega=0) $ no longer gives a diverging LR 
density-density response function. In particular,
since $V_C /(\sqrt{A^2(q) -1}) \to \infty$ for $q\to 0$,
the compressibility always vanishes as it should in a Coulomb
gas and a PS instability is now ruled out.
However, some finite-$q$ instabilities are still possible in the system
when 
\begin{equation}
P^{SR}(q,\omega=0)= -N(\sqrt{A^2(q) -1})/V_C \label{cdwcondition}
\end{equation}
leading
to a divergent $P^{LR}$. This is possible in principle, since
inside the PS region for the model without LRC forces,
$P^{SR}$ has a simple pole. Therefore $P^{SR}$ has a negative
branch from zero momentum up to sizable momenta 
(if the parameters are
choosen to be inside enough into the unstable region) 
and the condition (\ref{cdwcondition}) can be satisfied.
At this point, then, an instability occurs despite the
stabilizing effect of the LRC potential and an incommensurate
charge-density-wave (CDW) phase takes place in the system.

The $g$ vs $\delta$ phase diagram for various values of $V_C$ is
 reported in Fig. 8, where the continuous line 
represents the place at which $P^{LR}(q=q_c,\omega=0)$ 
diverges separating the stable uniform region from the 
unstable region where an incommensurate CDW is expected to form.
It is important to notice that, when the settling of a CDW
phase is given by a second order quantum transition,
no Maxwell construction is
needed to determine the stability region.

\begin{figure}
{{\psfig{figure=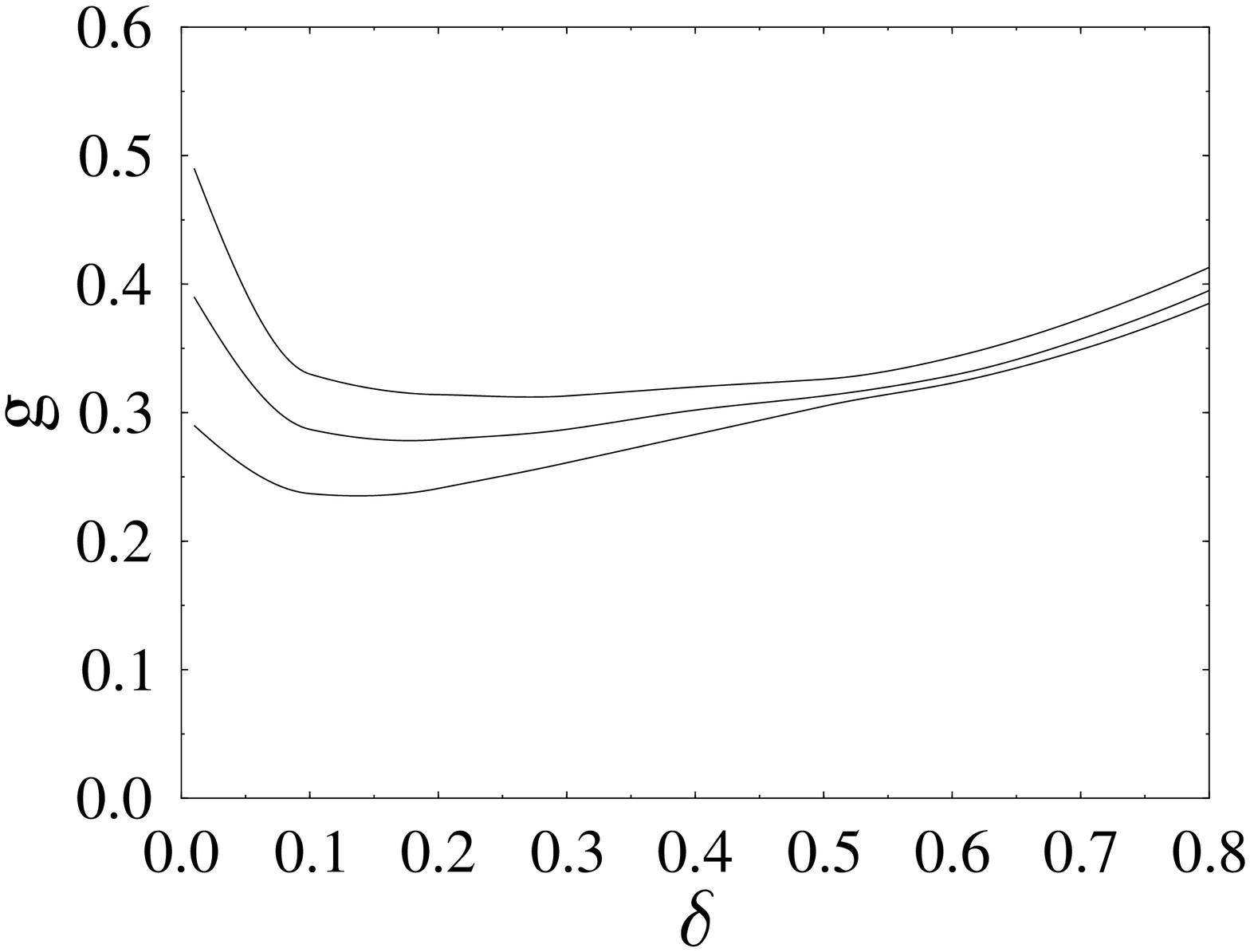,width=5.5cm,angle=-90}}}
{\small Fig 8. Phase diagram $e$-$ph$ coupling $g$ vs doping $\delta$
with $t_{\rm phys}=0.5$eV, $t'=-(1/6)t$, 
$\omega_{0\, \rm phys}=0.04$eV, and in the presence of
LRC forces with $V_{C\, \rm phys}=0.55$eV (lowest curve), 
$V_{C \, \rm phys}=1.65$eV (middle curve) and $V_{C \, \rm phys}=3.3$eV
(upper curve).}
\end{figure}
As it can be seen the effect of LRC forces is stronger at low doping.
This is so because in this region the poles of $P^{SR}$ tend to
occur at low momenta and are more effectively stabilized by the
Coulombic potential.

We found that the momenta $q_c$ at which 
the divergencies in $P^{LR}$ occur
obviously depend on the point $g$ vs $\delta$ and on the
strength of the Coulomb force $V_C$, but are generically 
sizable indicating that the wavelength of the
expected CDW phase is of a few unit cells. 

We performed an extensive analysis of the finite-$q$
instabilities for various values of doping, $t'/t$,
$g$, and $V_C$. It turns out that, for reasonable
values of these parameters, the instabilities
always occur at or close to the (1,0) and (0,1) directions.
This effect is due to 
the momentum structure of the short-range density-density
response function, which  is enhanced by the large
density of states in the (1,0) and (0,1) directions.
 
It is interesting to notice that, close to the finite-$q$
instability, the static effective scattering amplitude
in the particle-hole as well as in the particle-particle
channel diverge at finite momentum transfer like\cite{CDGPRL}
\begin{equation}
\Gamma(q,\omega=0) =- {A \over D' +B' \vert {\mathbf q}-{\mathbf q}_c
\vert^2}
\end{equation}
with $D'\propto (\delta-\delta_c)$ being a mass linearly vanishing by
approaching $\delta_c$.
The evolution
of $\Gamma^C(k_F,k_F',\omega =0)$ in approaching the unstable region
by doping variations 
is displayed in Fig. 9.
\begin{figure}
{{\psfig{figure=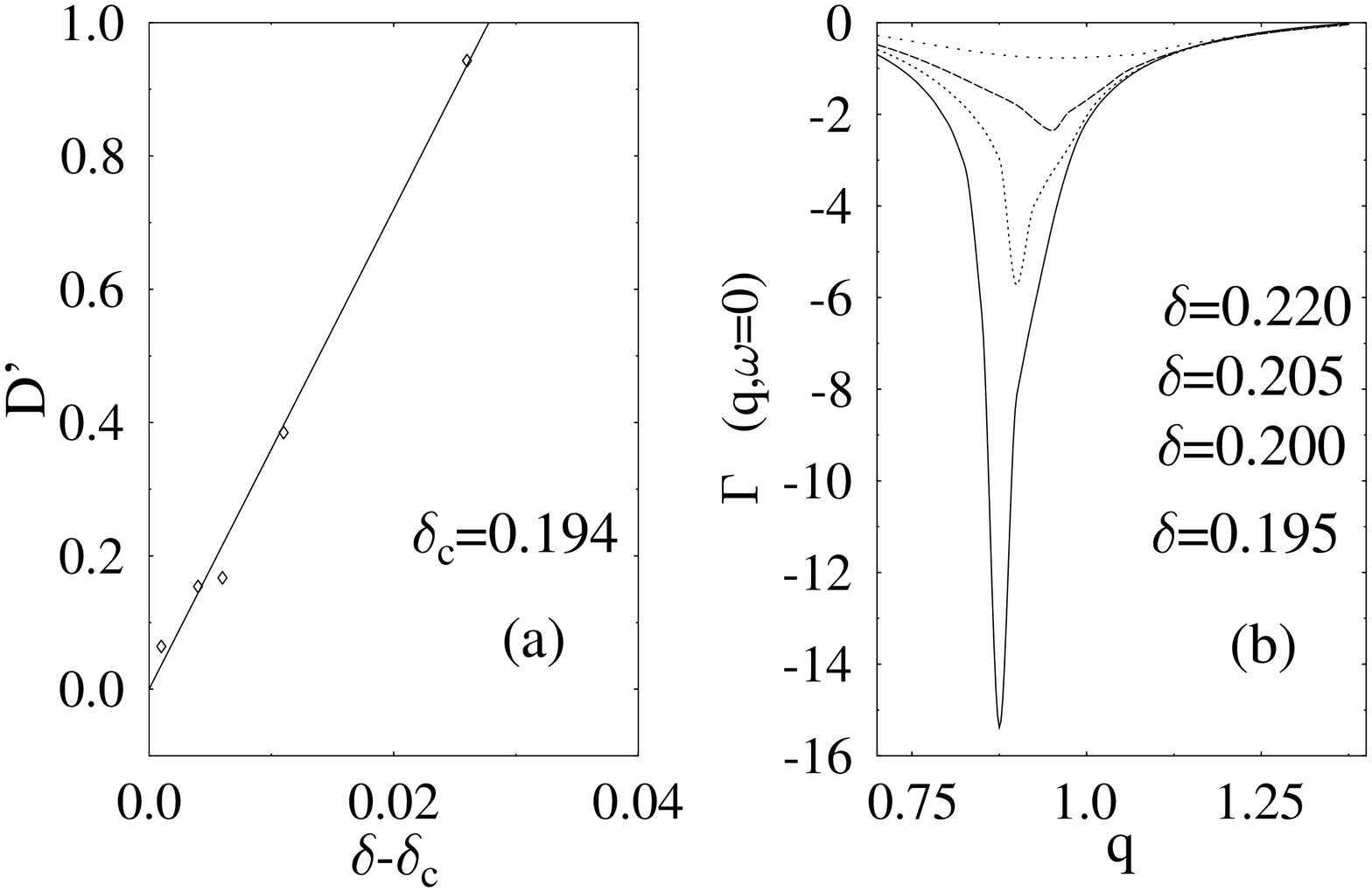,width=5.5cm,angle=-90}}}
{\small Fig 9. (a) Mass $D'$ of the effective static scattering amplitude
as a function of $\delta-\delta_c$ 
for $t_{\rm phys}=0.5$eV $t'=-1/6t$, $V_{C \, \rm phys}=0.55$eV, 
$\omega_{0\rm phys}=0.04$eV and
$g_{\rm phys}=0.240$eV. 
The open squares indicate the values of $D$ at various
dopings and the streight line is a linear fit. 
(b) Static scattering amplitude for the same parameters as in (a)
as a function of the transferred
momentum ${\mbox{\boldmath $q$}}$ in the ${\mbox{\boldmath $q$}}_c
\approx(\pm 0.28/a,\pm 0.86/a)$, direction. 
The doping $\delta=0.195,0.2, 0.205, 0.22$ increases
from the lower solid line to the upper dotted line.}
\end{figure}
Fig. 10 displays $\Gamma^C(k_F,k_F',\omega=0)$ over a large portion
of the Brillouin zone for parameter values close to those of a 
finite-q instability. It is evident that a deep
attractive interaction between the quasiparticles
arises on a broad momentum region. 
\begin{figure}
{{\psfig{figure=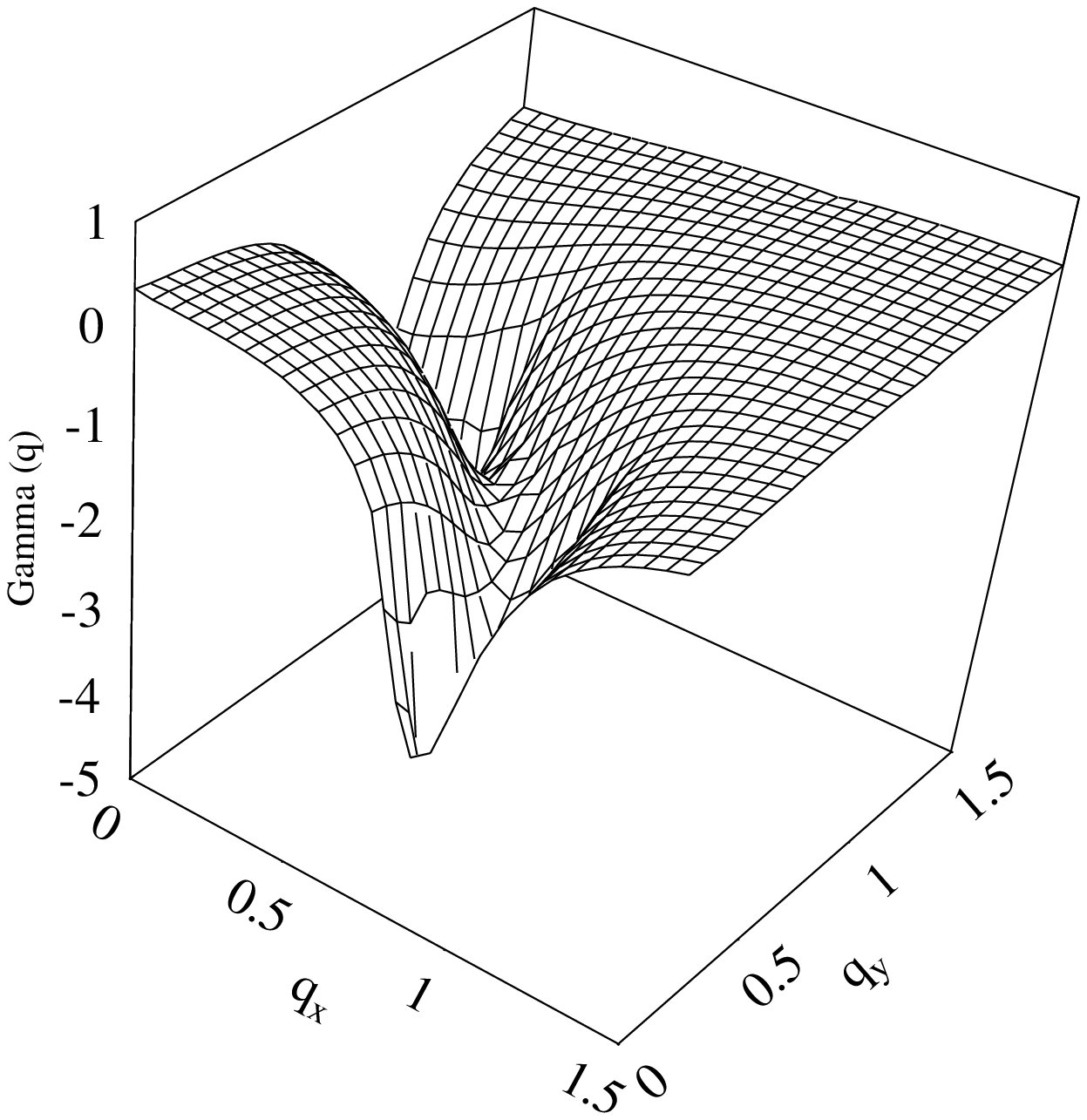,width=5.5cm,angle=-90}}}
{\small Fig 10. Momentum dependence of the static 
scattering amplitude for the same parameters as in Fig. 9 
at $\delta=0.195$.}
\end{figure}
This fact is particularly
remarkable in relation to the occurrence of Cooper instabilities.
As seen in Subsection III(a), a static Cooper instability
takes place when the Fermi-surface average of 
$-\Gamma^C(k_F,k_F',\omega =0)$ becomes positive [cf Eq.(\ref{coupl})]
and this seems likely to occur close to
finite-momentum instabilities.
The resulting coupling constants are reported in Table II.
\begin{table}
 \begin{center}
  \begin{tabular}{|c|c|c|c|c|c|c|}
   \hline
  $\delta$        & 0.300 & 0.305  & 0.310 & 0.330 & 0.360 & 0.400 \\
  $\lambda_{s_1}$ & 0.134 & 0.076 & 0.048 &-0.024  &       &       \\
  $\lambda_{d_1}$ & 0.372 & 0.206 & 0.174 & 0.110  & 0.070 & 0.045 \\
   \hline
   \end{tabular}
  \end{center}
{\small Extended $s$- and $d$-wave superconducting couplings 
for $t_{\rm phys}=0.5$eV, $t=-1/6t'$, $g_{\rm phys}=0.260/\sqrt{2}$eV,
$\omega_{0\rm phys}=0.04$eV and $V_{C\, \rm phys}=0.55$eV
at various dopings. The instability
is at $\delta_c=0.299$.}
\end{table}   
As expected, positive (i.e. attractive) coupling constants
arise close to the incommensurate CDW instability once more
supporting the idea that high-temperature superconductivity
could arise in the proximity of a charge instability. It is
also worth emphasizing that, like in the proximity of PS,
such a pairing instability takes place as an effect of
a q-independent phonon-induced attraction, which, close
to the CDW instability becomes highly structured in momentum
space, thereby opening the way to pairing symmetries other
than the simple $s$-wave. This fact accounts for the
attractive couplings also found in the $d$-wave channel. 
 In particular it turns out
that the $d$-wave symmetry of the order parameter is able
to take good advantage of the strong small-$q$ attraction 
and of the local (large-$q$) repulsion. 

In the region of small-intermediate $g$'s and dopings  
our analysis indicates the sure
existence of $d$-wave pairing in sizable  regions 
near the instability, whereas
the occurrence of $s$-wave pairing takes place in a much
narrower region. However, it should be
emphasized that the presence of a  $s$-wave static Cooper
instability  only in a narrow region, by no means excludes the
possibility of having $s$-wave superconductivity in a much larger
area of our phase diagram.  An appropriate   Eliashberg
dynamical analysis would be required to draw 
a firm conclusion, specially in the
light of the  results reported in the next section
showing a strong frequency dependence of the 
effective interaction between the quasiparticles.
Of course the same applies to the  attraction in the
$d$-wave channels, which could also be greatly favored by  dynamical
effects.

\subsection{Dynamical properties}
We carried out a dynamical analysis of the effective 
$e$-$ph$ vertex in the presence of LRC forces. 
The dynamical behavior of the $e$-$ph$ vertex in the presence 
of LRC forces is not very different from the behavior 
observed for $V_C=0$.
In both cases two different
screening regimes take place for small and large momenta.
Specifically,  when LRC forces are absent,
 we found that at small momenta the screening
energy $\omega_{scr}$ is proportional to the doping and
depends linearly on the exchanged momentum
$\omega_{scr}\approx v_F^* q$.
Instead, in the presence of LRC forces, it is found that
\begin{equation}
\omega_{scr}\approx \sqrt{v_F^* q}.
\end{equation}
The linear momentum dependence in the short-range-only
case is compatible with screening processes associated to 
both the particle-hole continuum and the zero sound.
In the presence of LRC forces, the square-root
momentum dependence is a clear indication that the plasmon
collective mode \cite{notaplasmon}
sets the cut-off energy at which small-momentum
screening processes cease to be relevant. 

On the other hand we found that at large momenta, a different
energy scale rules the screening processes, which,
both with and without LRC forces, is of the
order of the bare electron hopping $t$
$\omega_{scr}\approx \gamma t$.
This indicates that the local (large-$q$) physics
is governed by incoherent bare-electron processes
with typical energies of order $t$ much larger than the
typical energies of order $v^*_Fq \sim t\delta$ 
ruling the coherent quasiparticle processes.

The screening effects leading to different behaviors of the
effective $e$-$ph$ vertex at small and large momenta
also affect the dynamical scattering amplitude. This generic effect
is more evident in the proximity of an instability.
 Therefore in Fig. 11, we show the numerical evaluation
of the dynamical scattering amplitude.
A large attraction
at low frequency is found for transferred momenta close to
$q_c$ as a remnant of the infinite static attraction taking
place at the instability. 
\begin{figure}
{{\psfig{figure=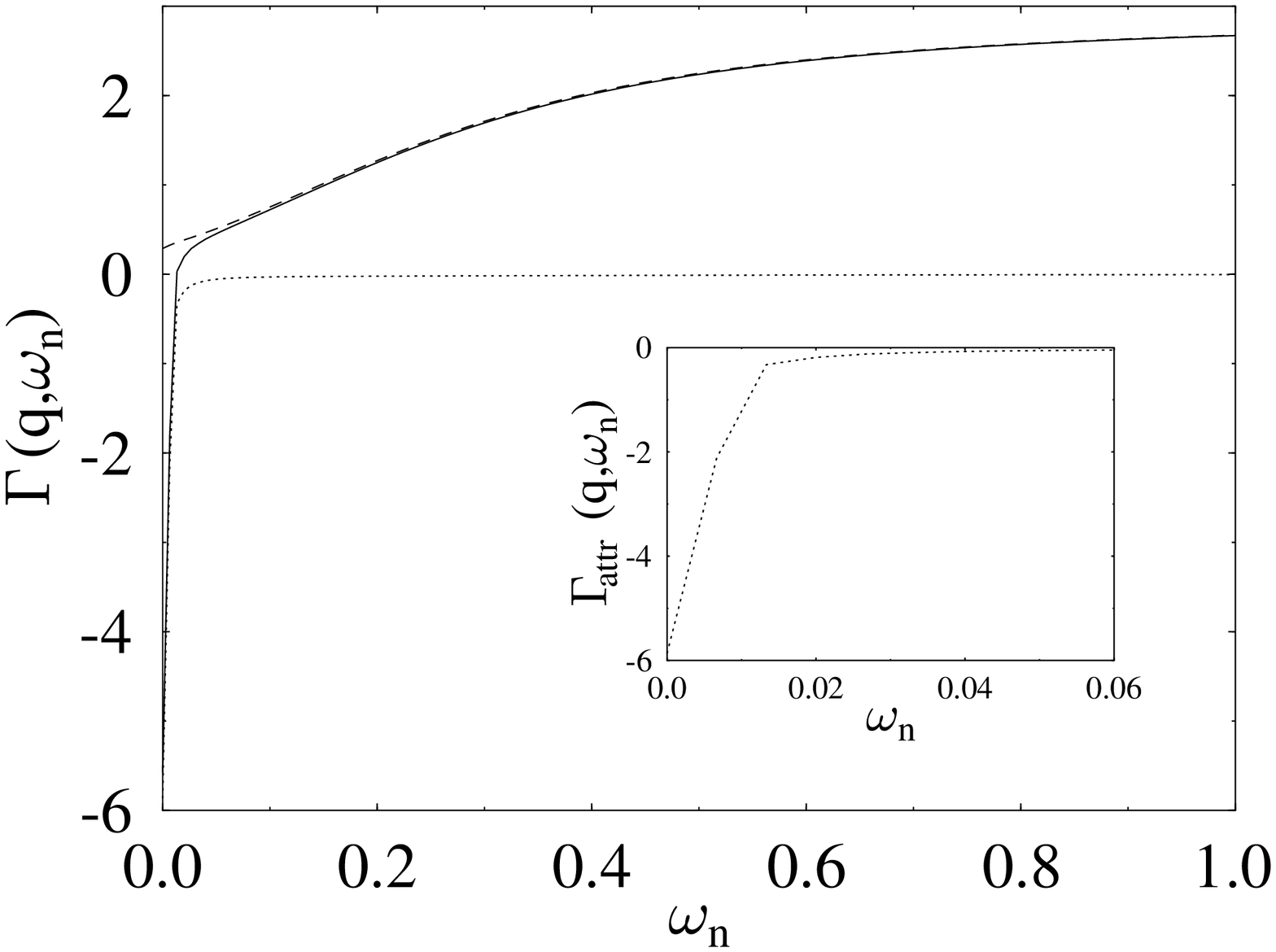,width=5.5cm,angle=-90}}}
{{\psfig{figure=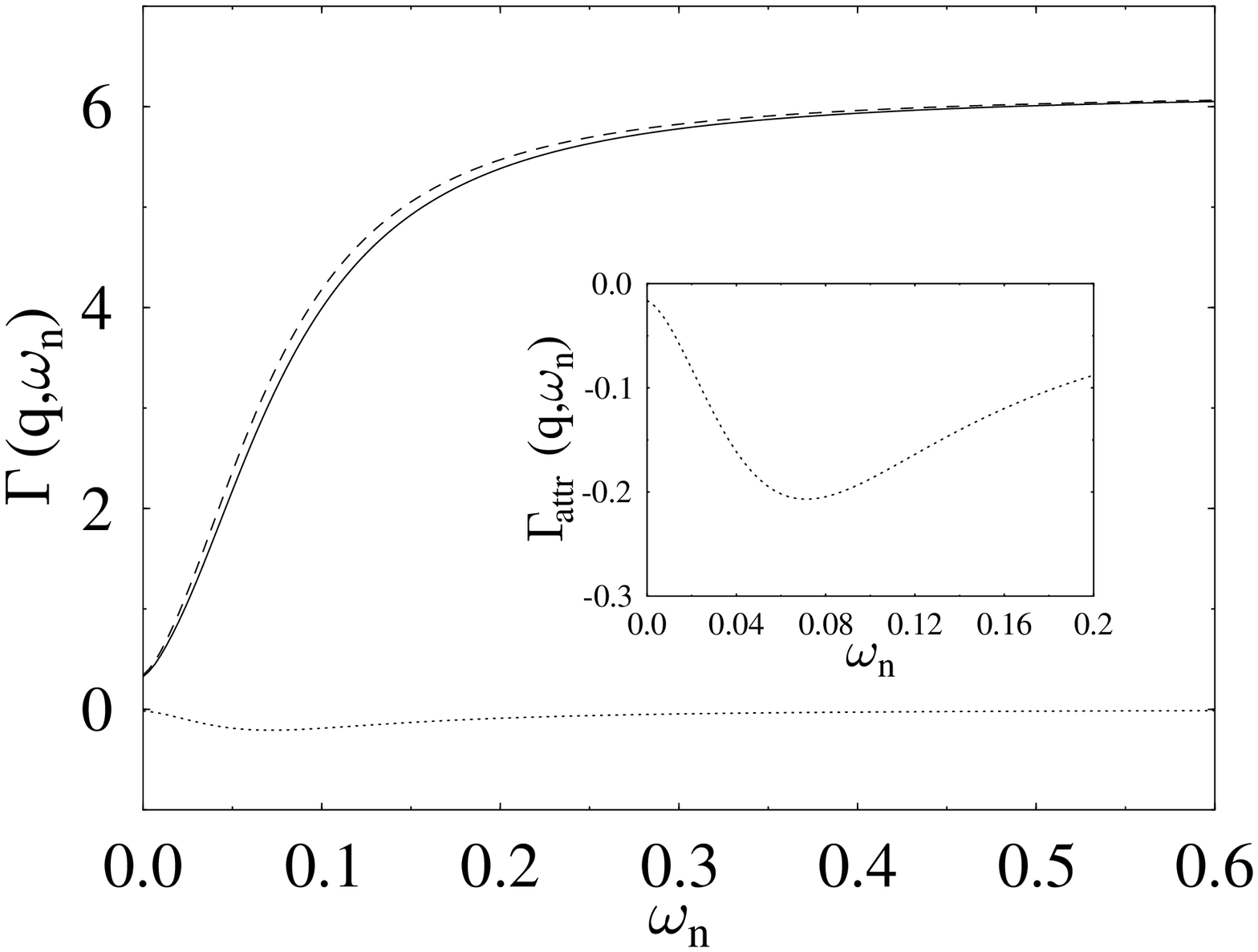,width=5.5cm,angle=-90}}}
{\small Fig 11. Effective scattering amplitude as a function of 
the transferred Matsubara frequency for 
$t_{\rm phys}=0.5$eV, $t'=-(1/6)t$, $g_{\rm phys}=
0.240/\sqrt{2}$eV, $\omega_{0\, \rm phys}=0.04$eV, and 
$V_{C\, \rm phys}=0.55$eV. The doping $\delta=0.2$ is close
to the critical value $\delta_c=0.195$.
Solid line: Total effective scattering amplitude $\Gamma$; dashed line:
Repulsive (slave- and Coulomb force bosons only) effective scattering
amplitude $\Gamma_{\rm rep}$; dotted lines: Attractive part
of the scattering amplitude $\Gamma_{\rm attr}=\Gamma - \Gamma_{\rm rep}$. 
The transferred momenta are in the
instability direction (0.28,0.86).
(a) Sizable transferred  momentum ${| \mbox{\boldmath $q$}|}=
{| \mbox{\boldmath $q$}_c|}=0.9$;
(b) small transferred momentum 
$|{\mbox{\boldmath $q$}}|=0.15$. The insets are enlargements of the
attractive parts of the effective scattering amplitudes.}
\end{figure}
This attraction is, however, rapidly
spoiled by increasing the Matsubara frequency. This is so
because at finite $q \approx q_c$ the dynamical screening
of the $e$-$ph$ vertex strongly suppresses the phononic
attractive part of the scattering amplitude. Therefore, as soon as 
the frequency spoils the static attraction due to the instability,
the attractive part of $\Gamma^C$ (dotted line in Figs. 11)
rapidly vanishes. An opposite behavior occurs at small momenta,
where no instability occurs and therefore
no static attraction is present. The usual $e$-$ph$
attraction only appears at a finite frequency around 
$\omega_0 > \omega_{scr} \approx q\delta $, where the $e$-$ph$
vertex is not severely suppressed. 

We conclude this subsection by recalling the result of Ref.\cite{CDGPRL}
where a real-frequency analysis was carried out finding strong
singular scattering between the quasiparticles of the form
\begin{equation}
\Gamma ({\mathbf q}, \omega)
\approx \tilde{U} -{A \over \omega_{\mathbf q}-i\omega}
\end{equation}
where $\omega_{\mathbf q} =D'+B' \vert {\mathbf q}-
{\mathbf q}_c \vert^2$
with $D'\propto (\delta-\delta_c)$. This scattering is of the form
proposed in Ref.\cite{pines} to explain the anomalous normal state 
properties of the superconducting copper oxides. Some substantial
differences are, however, worth being emphasized. First of all
the origin of the singular behavior is not necessarily
related to a magnetic scattering mechanism, in so far
it arises from the closeness to a charge instability. The 
mechanism driving the instability can be of various nature,
magnetic, excitonic or, as in the present model, phononic.
Secondly, as it can be seen in Fig. 10, a much more isotropic
region of large scattering arises in the present context
thereby bypassing the objection rised in Ref. \cite{hlublina}
that only a few ``hot'' points on the Fermi surface undergo
such strong scattering processes.

\subsection{The collective modes}
The analysis of the dynamical behavior of the model can be completed
by investigating the collective modes present in the system. This study
is particularly relevant for a deep understanding of the dynamical
mechanisms ruling the instability formation. 
To emphasize similarities and differences we consider
both the case with and without the LRC potential.

It was repeatedly pointed out in the context of models with short-range
interactions only, that PS occurs without a softening of a
massive collective mode. This was first established for the
charge-transfer mode in a three-band Hubbard model with NN 
Coulombic repulsion between copper and oxygen holes \cite{RCGBK}
and it was also confirmed in the case of a three-band Hubbard-Holstein
model with holes coupled to an optical phonon \cite{GC}.
A real-frequency analysis of both the imaginary 
part of the dynamical density-density correlation function
$Im \langle n n \rangle (q, \omega )$ and of the poles of
the phonon propagator $D^{a,a}(q,\omega)$ shows that also in 
the single-band infinite-U Hubbard-Holstein model, PS occurs
due to the phonon-mediated attraction between quasiparticles,
which pulls the zero-sound mode into the particle-hole continuum.
When the attraction is large enough, the velocity of this strongly damped
mode vanishes eventually driving the system unstable at zero momentum.
On the contrary, the phonon frequency, although sizably renormalized by the
$e$-$ph$ interaction stays finite all over the Brillouin zone. 

The scenario is strongly modified in the presence of LRC
forces, when the instability takes place at finite momenta.  
For such finite momenta, in the absence of $e$-$ph$ 
interaction the phonon is the mode at lower energy.  With the
introduction of a finite $e$-$ph$ coupling, it
turns out that the phonon is somewhat softened.
In particular the softening is complete,
i.e. the phonon energy vanishes  when the CDW instability
takes place. Again a comparative real-frequency analysis 
of $Im \langle n n \rangle (q, \omega )$ and of  
$ImD^{a,a}(q,\omega)$ allows to identify the nature of the
mode entering the continuum and producing the large
enhancement of the absorption at low-frequency.
Fig. 12(a) reports the behavior of the phonon frequency
[extracted from $ImD^{44}(q,\omega)$] as a function of
momenta in the specific direction at which the CDW
instability takes place [$(\pm 0.2, \pm 0.8)$ and 
$(\pm 0.8, \pm 0.2)$ for the parameters related
to 214 systems] at and above the critical doping
value. In Fig. 12(b) the phonon dispersion
are reported for the $(\pm 1, \pm 1)$ direction for the
same values of doping.
\begin{figure}
{{\psfig{figure=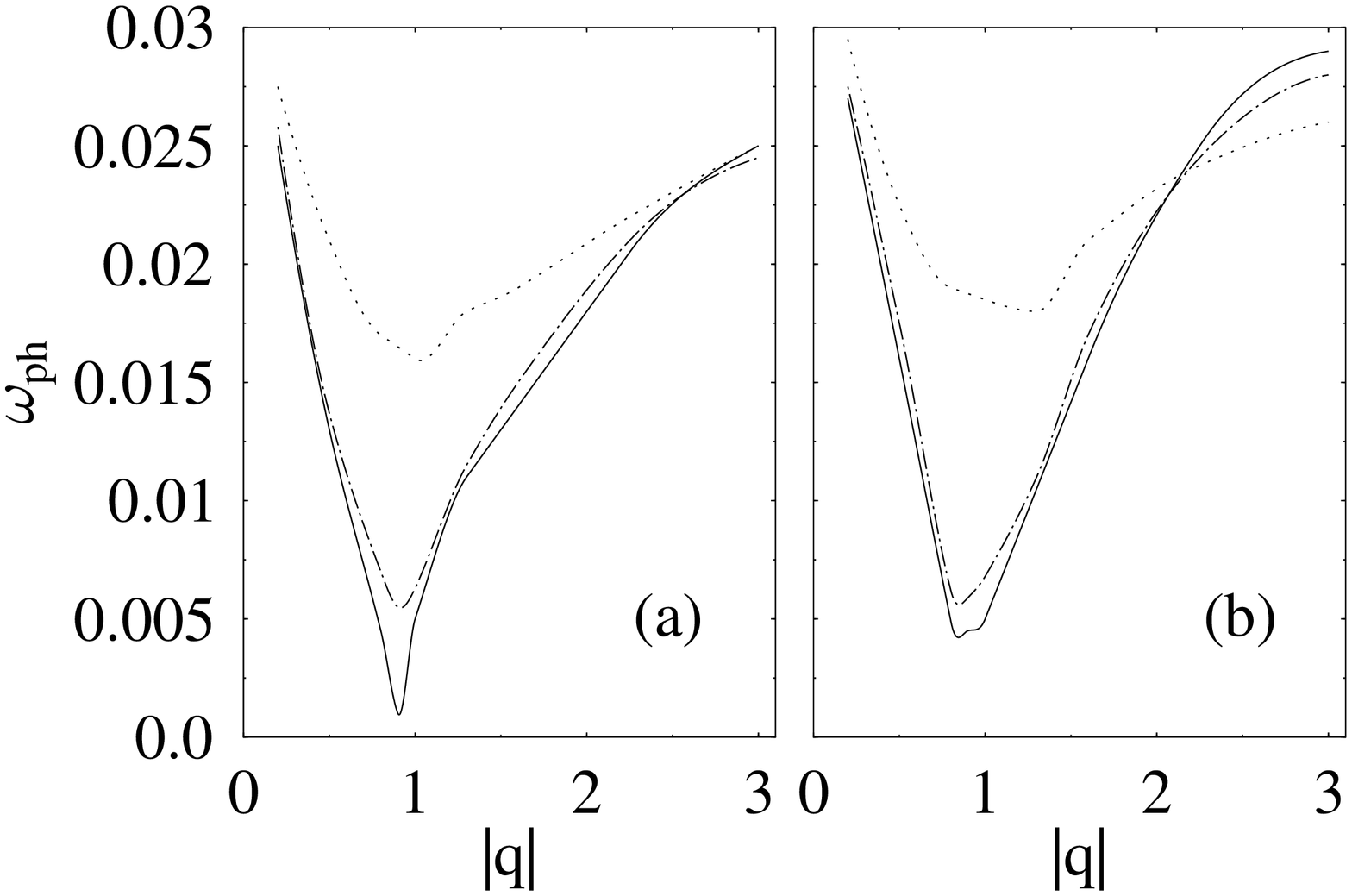,width=5.5cm,angle=-90}}}
{\small Fig 12. Phonon dispersion curves (a) in the 
instability (0.28,0.86) direction and (b) in the (1,1) direction, 
for $t_{\rm phys}=0.5$eV, $t'=-(1/6)t$, $g_{\rm phys}=
0.240/\sqrt{2}$eV, $\omega_{0\, \rm phys}=0.04$eV, and 
$V_{C \, \rm phys}=0.55$eV. The solid curves
correspond to the critical doping $\delta_c=0.195$, the 
dot-dashed and the dotted curves correspond to 
$\delta=0.21$ and $\delta=0.3$ respectively.}
\end{figure}
 Far from the critical momentum
the phonon frequency is sizably reduced with respect to
its bare value.  However, only a minor doping dependence of
this substantial reduction is visible in generic positions
in the Brillouin zone making the observation of the softening
effect rather difficult. On the other hand, a strong doping
dependence of the phonon dispersion is apparent in the proximity 
of the critical momentum.  In this momentum region the phonon 
completely softens at $\delta=\delta_c=0.195$. By increasing
the doping, i.e. by moving away from the instability
the softening becomes incomplete and progressively less
important although it seems to stay substantial up to $\delta=0.3$.

Some considerations are in order on the above analysis of the
collective modes in the proximity of charge instabilitites.
First of all a clear distinction can be made between the
behavior of the phonon mode in the absence and in the presence
of LRC forces. At a PS instability, the phonon always stays massive and
only mixes strongly with a low energy zero sound mode at 
finite but very small
momenta ($q \le \omega_0/v_F^*$). On the contrary the phonon
becomes completely soft at the CDW instability.
Therefore it would be natural to indicate neutron-scattering
experiments in copper oxides as crucial tests in order to
determine how close these material are to a {\it phonon-mediated}
CDW instability: if a substantial softening of some phonon mode
is found at some momenta, this would be a clear indication
of a frustrated PS almost leading to CDW.
Unfortunately, our schematic single-band Hubbard-Holstein
model is not detailed enough to provide indications of
the modes, which could undergo a visible softening
in real materials. Moreover, the strong reduction of the
phonon frequency is substantial on a sizable momentum range, but
is only found for doping  close to
the critical $\delta_c$, and this fact could render
the search for such an effect quite a difficult task.
An additional difficulty is that, due to the strong
mixing of electronic and phononic degrees of freedom,
the softened phonon modes would be greatly broadened
and they could well be seen only as an increase of 
spectral weight at low frequencies, quite differently
from the theoretical 1/N picture reported in Fig. 12.

\section{Discussion and conclusions}
In this paper we analyzed the screening processes 
and the occurrence of instabilities in the
Hubbard-Holstein model in the framework of Fermi-liquid
theory. In particular we investigated the role of
LRC forces in stabilizing PS and producing incommensurate
CDW instabilities. 

However, some limitations should be kept clear for a correct
understanding of the scenario here presented.
We carried out a leading order analysis  in 1/N, which
definitely neglects some effects, which are relevant
in a complete quantitative understanding of the real
materials. First of all our approach is designed to
deal with charge degrees of freedom, but lacks a correct
treatment of the spin degrees of freedom, which turn out to 
be crucial in the low-doping phase of the copper oxides.
Therefore antiferromagnetic correlations and the interplay
between spin and phonon degrees of freedom are absent in
our large-N, slave-boson model. These effects
only appear at higher order in 1/N. Secondly, our leading-order 
expansion does not allow for phonon vertex corrections
beyond the Migdal theorem, although in the low-doping region
the quasiparticle bands become very narrow thus leading to
a violation of the condition $E_F \gg \omega_0$. The absence
of these vertex corrections obviously rules out the possibility
of a correct observation of multi-phonon polaronic effects.

In the absence of
LRC forces, the scenario found here is consistent with previous
results obtained in the three-band Hubbard-Holstein model,
where PS was also found.
The main features observed here are i) a generic suppression
of the $e$-$ph$ vertex due to electronic screening also
resulting in a strong dependence on the $v_Fq/\omega$
ratio; ii) the persistent possibility of PS arising
from the phonon-mediated attraction; iii) 
a singular behaviour of the effective interaction
between quasiparticles at low momenta
close to the PS region. 

The main achievement of the present work is, however, the
analysis of the model in the presence of LRC forces. In particular
it is remarkable that, when LRC forces are
included in the model, PS is spoiled, but 
 finite momentum instabilities still
take place on substantial and quite physical regions of the
parameter space. Singular effective interactions 
are again obtained at finite momenta in the proximity of the
CDW instability;
iv) Cooper pairing both in the $s$- and the $d$-wave channels
is present already in the static limit for systems close
enough to the unstable regions, both in the presence and in the
absence of LRC forces.

The present model provides a rather simple playground to
investigate the generic properties of electronic
systems close to charge instabilities. Indeed, while some
properties like, e.g., the behavior of the phonon mode, 
are strictly related to the
phononic nature of the interaction here considered, the
main results obtained here are generic of models\cite{GRCDK1}
showing charge instabilities. In particular we believe that the strict
relation between charge instabilities, strong scattering
and Cooper pair formation [points iii) and iv) above] 
is a generic feature of strongly correlated electron systems
 irrespective of the underlying physical  mechanisms
leading to PS or to incommensurate CDW.

\centerline{\bf ACKNOWLEDGEMENTS}
We acknowledge interesting discussions with Prof. C. Castellani.
This work was partly supported by the Istituto Nazionale
per la Fisica della Materia (INFM).

\begin{appendix}
\section{The Maxwell construction}
In order to perform the Maxwell construction, we modify the model in Eq.
(\ref{HHHam})
coupling the phonons to the full electron density rather than to
the density fluctuations. In this way an $e$-$ph$ coupling is 
effective already at mean-field level, where a non-zero mean-field
value of the phonon field $a_0=\langle a \rangle =\langle a^\dagger \rangle$
arises. The mean-field Hamiltonian acquires the form
\begin{eqnarray}
H_{MF}'& =  &  \sum_{k \sigma } E_k
c^{\dagger}_{k \sigma } c_{k\sigma} -(\mu_0 -\lambda_0 
+2ga_0) \sum_{k \sigma }
c^{\dagger}_{k \sigma } c_{k\sigma} \nonumber \\
& + & N\lambda_0 \left( r_0^2-{1 \over 2}\right) +N\omega_0 a_0^2 
\end{eqnarray}
Minimizing the mean-field free energy with respect to $a_0$ 
allows to obtain the self-consistency equation for $a_0$
\begin{equation}
N\omega_0a_0=Ng\sum_k f\left( E(k) \right)=
g{N\over 2}\left(1-\delta \right)
\end{equation}
This determines the phonon-induced shift of the chemical potential
giving rise to a doping dependent correction to this quantity. 
Owing to the standard expression for the compressibility
$\kappa\equiv {\partial n \over \partial \mu}=
-{\partial \delta \over \partial \mu}$, the
$q=0$ instability can be read directly from the $\mu$ vs $n$
curves: a stationary point in $\mu (\delta)$ corresponds to an
infinite compressibility and a standard Maxwell construction
in $\mu (\delta)$ determines the region where the system is in a 
single phase.

\end{appendix}

%%%%%%%%%%%%%%%%%%%%%%%%%%%%%%%%%%%%%%%%%%%%%%%%


\begin{references}
\bibitem{notaRCGBK} This second possibility was suggested for 
different model in Ref. \cite{RCGBK}
together with the alternative possibility of bubble formation
of hole-rich and  hole-poor phases.
\bibitem{notaeqil}A different possible approach is also viable,
   where the phonons are coupled with the density and not with its
fluctuations only. Then the ion equilibrium positions must be 
determined self-consistently together with the other
mean-field parameters as a function of doping and of the
tight-binding parameters. In this latter scheme the effect of
phonons is already present at mean-field level and plays a role
in the behavior of the chemical potential as a function of filling.
Therefore this scheme is needed in order to apply
the Maxwell construction to determine the phase-separation
 region. The approache described in the text 
 assumes instead that the mean-field effect of phonons is
included in the bare tight-binding parameters. Despite
this seeming difference, we checked that the
two approaches give the same results (see Sec. III).
\bibitem{notazeyher}In Ref.\cite{zeyher}, however, only the case
with $\alpha=t'/t=0$ was considered, which is rather peculiar.
This can be seen by considering again the small-$q$ case
of Eq.(\ref{wi2}) for $t'=0$. In this particular case 
the effect of the static vertex at 
low doping is much less dramatic since this vertex is proportional 
to $\left(1+4tN\nu^*\varepsilon_{k_F}(\alpha=0) \right)^{-1}$, where 
$\varepsilon_{k_F}(\alpha=0)=(\cos (k_{Fx})+\cos (k_{Fy}))$ vanishes 
when $\delta \to 0$.
This peculiarity of the single-band Hubbard 
model is not present once next-NN
hopping terms are considered. In this latter case
$\varepsilon_{k_F}$ no longer vanishes at half-filling 
and the presence of a large quasiparticle density of 
states $\nu^*$ in the denominator 
leads to a large suppression of the e-ph interaction.
\bibitem{notacdwsr}On the other hand it turns out 
that for unrealistically large doping the instability 
first takes place at sizable finite momenta (dashed line)
leading to the formation of
incommensurate charge-density waves (CDW). In particular we found that
the finite-momentum instability first takes place for momenta in the 
$(\pm 1,0)$ and $(0,\pm 1)$ directions and momenta $q=2k_F$ connecting
regions with slightly larger density of states.
\bibitem{notacdwsrb}Attractive superconducting couplings are also found
both in the $s$-vave and the $d$-wave channel in the region
where the instability takes place at finite momenta.
\bibitem{ef} $\varepsilon_{k_F}$
vanishes when $\alpha =t'/t =0$ and $\delta \to 0$.
This effect simply arises from the specific form of the Fermi
surfaces in the single-band Hubbard model with NN hopping only, 
where the constant-energy surfaces are determined by the condition 
$\cos (k_{xF})+\cos (k_{yF})={\rm{constant}}$. Then, 
at $\delta=0$, the Fermi energy reduces to a square
where the constant is zero. Therefore, when the next-NN 
hopping integral $t'$ is zero, one finds that,
for any finite $e$-$ph$ coupling $g$  the attractive
part will dominate. At the same time the quasiparticle density
of states diverges ($\nu^* \approx 1/\delta$) so that  a very large
and negative $F_0^s$ arises. This shows that, at least within our
large-$N$ scheme, a phase-separation instability 
is always obtained for any finite $g$ 
when the doping is small enough and $t'=0$.
This is no longer true when a finite next-NN hopping integral $t'$
is considered, which modifies the shape of the Fermi surface
and leads to non-zero values for $\varepsilon_{k_F}$ at zero doping.
\bibitem{notaextleg}It should be noted that the quasiparticle-boson
vertices [Eqs.(\ref{lamr})-(\ref{lamg})] are constant or only depend
on the incoming fermionic momenta via expressions
proportional to the energy. As a result  the
quasiparticle on the Fermi surface only couple to the bosons
via constant vertices $\Lambda_r =  -4tr_0^2 \varepsilon_{k_F},
\Lambda_\lambda =  i$ 
and $\Lambda_a =-2g$, irrespective on their momenta.
\bibitem{notacccdgr}This possibility was shown within
a mean field treatment of the Kondo
lattice model in Ref.\cite{CCCDGR}.
\bibitem{notafl}These conditions are not very restrictive and are
    realized, e.g., in the important case of strongly interacting
    fermions close to a metal-Mott-Hubbard or
    Charge-Transfer insulator transition.
\bibitem{notaplasmon}It should be remembered that in twodimensional
systems embedded in a  threedimensional space, the plasma modes
have a square-root momentum dependence, at least at small momenta.
\end{references}
\end{document}